\documentclass[10pt,aps,pra,twocolumn,superscriptaddress,reprint,amsmath,amssymb,showkeys,floatfix]{revtex4-2}

\usepackage{graphicx}       
\usepackage{mathtools}
\usepackage{tcolorbox}     
\usepackage{physics}
\usepackage{relsize}

\newcommand{\e}{\mathrm{e}}
\newcommand{\til}{~}
\usepackage[ colorlinks=true, linkcolor=black, 
  citecolor=blue,
  urlcolor=blue
]{hyperref}

\begin{document}

\title{Mass-imbalanced two-dimensional Bose-Fermi mixtures with boson-fermion pairing}

\author{Cristiano L. K. Chiarappa}
\affiliation{Dipartimento di Fisica e Astronomia \textquotedblleft Augusto Righi", Università di Bologna, Via Irnerio 46, I-40126, Bologna, Italy}

\author{Pietro Bovini}
\author{Pierbiagio Pieri}
\email{pierbiagio.pieri@unibo.it}

\affiliation{Dipartimento di Fisica e Astronomia \textquotedblleft Augusto Righi", Università di Bologna, Via Irnerio 46, I-40126, Bologna, Italy}
\affiliation{INFN, Sezione di Bologna, Viale Berti Pichat 6/2, I-40127, Bologna, Italy}

\date{\today}

\begin{abstract}

We analyze a two-dimensional Bose-Fermi mixture at zero temperature in the presence of a tunable Bose-Fermi attraction. We adopt a diagrammatic $T$-matrix approach and study the behavior of several thermodynamic quantities for the two species as functions of density, mass ratio, and coupling strength. These include the chemical potentials, the boson momentum distribution function, the condensate density, and Tan’s contact parameter. We analytically demonstrate that the present $T$-matrix formalism recovers the correct second-order perturbative expansion of the chemical potentials in the weak-coupling regime, and test it numerically. The near-universal behavior of the condensate fraction already found in prior work for the mass-balanced case is confirmed for different masses and becomes even more accurate when the boson mass is large. The mass imbalance emerges as an additional control parameter that qualitatively affects the bosonic momentum distribution. 
In particular, we found that it can be used to allow for the experimental observation of a peculiar peak in the boson momentum distribution at finite momentum.

\end{abstract}

\maketitle

\section{\label{sec:intro}Introduction}

Ultracold quantum gases provide a controlled environment that exhibits genuinely quantum-mechanical behavior. Among these, Bose-Fermi (BF) mixtures have been studied extensively both theoretically \cite{Sachdev-2005,Stoof-2005,Storozhenko-2005,Avdeenkov-2006,Pollet-2006,Rothel-2007,Barillier-2008,Pollet-2008,Bortolotti-2008,Parish-2008,Watanabe-2008,Fratini-2010,Zhai-2011,Song-2011,Ludwig-2011,Fratini-2012,Anders-2012,Bertaina-2013,Fratini-2013,Sogo-2013,Guidini-2014,Guidini-2015,Bertaina-2015,Kharga-2017,Ohashi-2019,Manabe-2020,Ohashi-2021,Schmidt-2022,Tajima-2024,Shen-2024,Pastukhov-2024,2dbfmix,Gualerzi-2025,Foster-2026} and experimentally \cite{Jin-2008,Zwierlein-2011,Zwierlein-2012,Zwierlein-2012a,Ketterle-2012,Cumby-2013,Jin-2013a,Porto-2015,Salomon-2014,Salomon-2015,Jin-2016,Pan-2017,Gupta-2017,DeSalvo-2017,Ferlaino-2018,Grimm-2018,Patel-2023-2,Valtolina-2019,Zwierlein-2020,Khoon-2020,Ozeri-2020,Fritsche-2021,Schindewolf-2022,Bloch-2022,Bloch-2023,Patel-2023,Baroni-2024,Yan-2024,flty-9d72,DeMartino-2025,Ando-2025,Grimm-2019}, as their high degree of tunability enables the exploration of diverse many-body effects.
Optical or magnetic trapping techniques make it possible to cool atomic vapors to the nanokelvin scale and reach the quantum degenerate regime, in which bosons and fermions exhibit their characteristic quantum statistics \cite{Porto-2015, Ferlaino-2018, Pan-2017, Khoon-2020}. 
In addition, Feshbach resonances offer precise control over interparticle interactions, while the trapping geometry controls the atomic density distribution.
These techniques also allow for the tuning of the system's effective dimension. In particular, confinement-induced resonances provide access to a dimensional crossover from a 3D confinement down to a (quasi-)2D system.

The realization of BF mixtures has allowed exploration of a wide range of phenomena like phase separation \cite{Grimm-2018}, polarons \cite{Fritsche-2021,Jin-2016,Schmidt-2018,Zwierlein-2020,Scazza-2022,Massignan-2014}, dual superfluidity \cite{Salomon-2014,Gupta-2017,Wu-2018,Salomon-2015}, collective excitations \cite{Patel-2023,Yan-2024}, mediated interactions \cite{Bruun-2018,Ozeri-2020,Shen-2024, Baroni-2024}, and Feshbach molecules \cite{Jin-2008,Jin-2013a,Zwierlein-2012,Ketterle-2012,Valtolina-2019,Schindewolf-2022,Bloch-2023,Cumby-2013,Patel-2023-2,flty-9d72}, mostly in three-dimensional ultracold atomic mixtures. 

However, two-dimensional systems are expected to attract considerable experimental interest for several reasons.
First, a combination of a Feshbach resonance and a confinement-induced resonance allows the simultaneous tuning of both the BF attraction and the boson-boson (BB) repulsion \cite{Petrov-2001,Parish-2015,Haller-2010}, with the latter being crucial for ensuring the mechanical stability of a homogeneous mixture. 
Moreover, recent theoretical work \cite{PhysRevLett.121.263001} has shown that weakly bound BF molecules in the presence of a finite BB repulsion are collisionally stable and are predicted to exhibit strong $p$-wave pairing, potentially leading to $p$-wave superfluidity. 
The realization of $p$-wave superfluidity could open the way to the exploration of topologically protected states, including Majorana modes and non-Abelian vortex excitations \cite{Gurarie-2007,Dalibard-2019}, thereby providing a potential route toward fault-tolerant quantum computation \cite{Kitaev-2001,Kitaev-2003,Lian-2018,Aasen-2016,Sarma-2015,Knapp-2018}. 

The quest for $p$-wave superfluidity is actively pursued also in BF mixtures  realized in semiconducting nanostructures belonging to the class of transition metal dichalcogenides (TMDs) \cite{Schmidt-2023, Zerba-2023}. TMDs are layered systems consisting of crystalline sheets of transition-metal and chalcogen atoms \cite{RevModPhys.90.021001,Imamoglu-2021,Jauregui-2019,Fey-2020,Kuhlenkamp-2022,PhysRevLett.105.136805,Sidler-2017,Bruun-2021-PRL,kang-2023,excpolbec2,Manca2017,Glazov_2021,Fogler2014,adlong-2025}. Newly developed exfoliation techniques have enabled the fabrication of few- and mono-layer TMD structures, in which charge carriers are effectively confined to two dimensions.
In their monolayer form, TMDs can exhibit a direct band gap \cite{PhysRevLett.105.136805}, allowing efficient optical excitation across the gap. Such excitations create an electron in the conduction band and a hole in the valence band, which interact through the Coulomb force and may form a bound state known as an exciton.
Excitonic states in TMD monolayers are more strongly bound compared to those in bulk TMD semiconductors due to the enhancement of Coulomb attraction in reduced dimensions. Their typical binding energy is of the order of hundreds of meV \cite{Fey-2020,excpolbec2}, often exceeding the electron Fermi energy by an order of magnitude for not too high a density of charge carriers. If these remain sufficiently dilute compared to the spatial extent of excitons, the latter can then be treated effectively as atom-like bosons even in the presence of the fermionic medium. This allows for the investigation of mixtures of excitons and doped charges in analogy with atomic BF mixtures.

Excitons can be involved in bosonic scattering processes \cite{Manca2017} and are expected to exhibit collective phenomena such as Bose-Einstein condensation and superfluidity \cite{Glazov_2021,Fogler2014}. Regarding the BF interactions, an electrically tunable Feshbach resonance has recently been realized in a MoSe$_2$ bilayer system, enabling precise electrical control over the exciton-hole interaction strength \cite{Imamoglu-2021}. This electrical tunability offers a promising route toward realizing degenerate BF mixtures with fully controllable BF interactions in a purely two-dimensional solid-state platform.

In a mixture of excitons and doped charges, the bosonic and fermionic components are mass-imbalanced, with the exciton effective mass being roughly twice that of the charge carriers in typical TMDs \cite{adlong-2025,Fey-2020}. A similar imbalance arises naturally in ultracold atomic mixtures, where it can vary substantially depending on the species involved. In Table \ref{tab:bf-mixtures}, we report a list of experimentally realized ultracold heteronuclear BF mixtures, their mass ratios, and the corresponding references. The wide range of mass-imbalance values between the bosonic and fermionic species, appearing in both ultracold atomic gases and TMDs, motivates the present work.

We extend a previous analysis \cite{2dbfmix}, which considered a mass-balanced two-dimensional Bose–Fermi mixture with attractive BF interactions at zero temperature, to the experimentally relevant case of mass-imbalanced mixtures.

Our main findings can be summarized as follows. The mass imbalance produces significant quantitative modifications of thermodynamic quantities such as the chemical potentials, while leaving their qualitative behavior largely unchanged. By contrast, the bosonic momentum distribution exhibits qualitative changes. In particular, we identify a range of parameters for which a predicted finite-momentum peak should remain observable in a mechanically stable system. 

The near-universal behavior of the condensate fraction found in \cite{2dbfmix} is affected by the mass ratio, with universality becoming more accurate for larger boson masses. 
Finally, the hybridization between molecular and atomic states previously found in \cite{2dbfmix} 
is significantly affected by the mass ratio, with large boson masses suppressing it and thus making the dispersion of dressed BF pairs closer to its not-hybridized form.

The article is organized as follows. 

In Sec.\til\ref{sec:methods}, we illustrate the system Hamiltonian, explain our choice of diagrams, and show the set of thermodynamic equations to be solved.
Section\til\ref{sec:xuno} reports the numerical results for chemical potentials, Tan's contact parameter, and boson momentum distribution in a mixture with an equal number of bosons and fermions for several values of mass ratios.
In Sec.\til\ref{ximnoreuno} we report the boson momentum distribution for a system with a majority of fermions and study the effects of mass imbalance on the condensate density and the hybridization energy scale introduced in \cite{2dbfmix}. 
Finally, in Sec.\til\ref{conclusions} we draw our conclusions. The two appendices provide details on the weak- and strong-coupling expansions of our theory.
\begin{table}[t]
\centering
\begin{tabular}{|c|c|c|}
\hline
{Mixture (B$\,$-$\,$F)} & \textbf{$m_\mathrm{B}/m_\mathrm{F}$} & {References} \\
\hline
& & \\
$^{87}$Rb$\,$-$^{171}$Yb & 0.509 & \cite{Porto-2015} \\
$^{23}$Na$\,$-$^{40}$K & 0.575 & \cite{flty-9d72,  Zwierlein-2012a, Zwierlein-2012,Bloch-2023,Schindewolf-2022,Zwierlein-2020,Yan-2024} \\
$^{168}$Er$\,$-$^{161}$Dy & 1.043 & \cite{Ferlaino-2018}\\
$^{87}$Rb$\,$-$^{40}$K & 2.175 & \cite{Jin-2013a,Cumby-2013,Jin-2008,Valtolina-2019,Jin-2016,Ozeri-2020} \\
$^{23}$Na$\,$-$^{6}$Li & 3.833 & \cite{Ketterle-2012} \\
$^{41}$K$\,$-$^{6}$Li & 6.833 & \cite{Pan-2017,Grimm-2019,Fritsche-2021,Wu-2018,Grimm-2018,Baroni-2024}\\
$^{84}$Sr$\,$-$^{6}$Li & 14 & \cite{Khoon-2020}\\
$^{133}$Cs$\,$-$^{6}$Li & 22.167 & \cite{Patel-2023-2,Patel-2023}\\
$^{174}$Yb$\,$-$^{6}$Li & 29 & \cite{Gupta-2017}\\
& & \\
\hline
\end{tabular}
\caption{Representative experimentally realized ultracold heteronuclear BF mixtures, their mass ratios, and the corresponding references.}
\label{tab:bf-mixtures}
\end{table}

\begin{figure*}[t]
\includegraphics[width=1.0\linewidth]{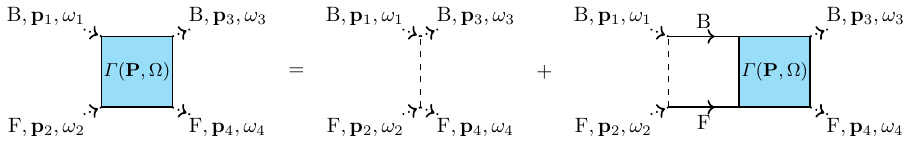}
         \caption{Feynman diagrams for the particle-particle ladder $\varGamma(\textbf{P},\Omega)$, corresponding to the $T$-matrix in the normal phase. Full lines correspond to bare bosonic (B) and fermionic (F) Green's functions, dashed lines correspond to bare BF interactions, and dotted lines with arrows indicate the external momentum flow. $(\textbf{P},\Omega) = (\textbf{p}_1+\textbf{p}_2, \omega_1+\omega_2)$ is the center of mass (2+1)-momentum.}
    \label{fig1}
\end{figure*}

\section{Methods}\label{sec:methods}

We consider a dilute homogeneous mixture of spinless bosons and spin-polarized fermions with interspecies attraction in a two-dimensional setting in the zero-temperature limit, in which bosons may condense even in 2D \cite{Hohen-1967}. In a previous work \cite{2dbfmix}, we studied the thermodynamic properties of the same system at
$T = 0$ within a $T$-matrix diagrammatic approach extended to the condensed phase. The strengths and limitations of this method are discussed in \cite{2dbfmix} and in the references cited therein.

Here, we are interested in the case of mass-imbalanced mixtures, which was not previously addressed. In particular, we will consider several possible values of the mass ratio parameter $\gamma_\mathrm{m} = m_\mathrm{B}/m_\mathrm{F}$ between the boson and fermion masses $m_\mathrm{B}$ and $m_\mathrm{F}$, respectively.
As in \cite{2dbfmix}, we focus on mixtures with bosonic density $n_\mathrm{B}$ not exceeding the fermionic density $n_\mathrm{F}$. In terms of the bosonic concentration parameter $x = n_\mathrm{B}/n_\mathrm{F}$, this corresponds to $ x\le 1$.

For a thorough description of the formalism, the reader is referred to \cite{2dbfmix}. In the following section we recap the main equations and quantities relevant to the present work.

We set $\hbar = 1$ throughout this paper.

\subsection{Formalism}

\subsubsection{Hamiltonian}

In the presence of a broad Feshbach resonance tuning the strength of the BF attractive interaction, the (grand-canonical) Hamiltonian is given by

\begin{align} 
       {\mathcal{H}} &= \mathcal{H}_{\rm BF}+V_{\rm BB} \nonumber\\
       &=\sum_{s = \mathrm{B},\mathrm{F}}\int \dd^2 r\; \psi_s^\dagger(\textbf{r})\left(-\frac{\nabla^2}{2m_s}-\mu_s\right)\psi_s(\textbf{r})\\&+v_0^\mathrm{BF}\int\dd^2 r\; \psi_\mathrm{B}^\dagger(\textbf{r}) \psi_\mathrm{F}^\dagger(\textbf{r}) \psi_\mathrm{F}(\textbf{r}) \psi_\mathrm{B}(\textbf{r})  + V_{\rm BB}, \label{eqn:hamiltonian}
\end{align}
where $\psi^{\dagger}_s$ and $\psi_s$ create and destroy respectively a particle of mass $m_s$ and chemical potential $\mu_s$, with $s=\mathrm{B,F}$ indicating the bosonic or fermionic nature. 

Here, $\mathcal{H}_{\rm BF}$ describes the kinetic energy and the interaction between bosons and fermions. At the same time, $V_{\rm BB}$ considers the (possible) presence of a (weak, non-resonant) boson-boson repulsion. 

The BF attractive contact interaction is regularized by expressing the bare strength $v_0^{\rm BF}$ in terms of the (2D) BF scattering length $a_{\rm BF}$ as
\begin{equation}\label{eqn:v0bf}
\frac{1}{v_0^\mathrm{BF}} = - \int \frac{\dd^2 k}{\left( 2 \pi \right)^2} \frac{1}{\epsilon_0+ {\textbf{k}^2}/{(2m_\mathrm{r})}},
\end{equation}
where $\epsilon_0 = 1/(2m_\mathrm{r}a_\mathrm{BF}^2)$ is the BF binding energy of the associated two-body problem in vacuum with $m_\mathrm{r} = m_\mathrm{B} m_\mathrm{F} / (m_\mathrm{B}+m_\mathrm{F})$ being the reduced mass of the BF pair.

The ultraviolet divergence in the integral on the right-hand side of Eq.\til\eqref{eqn:v0bf} compensates analogous divergences occurring in many-body diagrammatic perturbation
theory. For details on the regularization procedure, see, e.g., \cite{Miyake-1983} and the Supplemental Material of \cite{Bauer-2014}.

One can define an effective BF dimensionless coupling parameter $g = \ln\left[ 1/(k_{\rm F} a_{\rm BF}) \right]$ using the Fermi momentum $k_{\rm F}=\sqrt{4\pi n_{\rm F}}$ of a free Fermi gas with density $n_{\rm F}$, in analogy with 2D fermionic systems.

With the above definition of $g$, one can express the BF binding energy in units of the Fermi energy $E_{\rm F} = k_{\rm F}^2/(2 m_{\rm F})$ as $\epsilon_0/E_{\rm F}  =\e^{2g}(\gamma_\mathrm{m}+1)/\gamma_\mathrm{m}$. In terms of $g$, the weak- and strong-coupling limits correspond to $g \to -\infty$ and $g \to +\infty$, respectively. In practice, for the range of masses explored in the present work, they are well represented by $g \lesssim -2$ and $g \gtrsim  2$.

\subsubsection{Many body \texorpdfstring{$T$-matrix}{T-matrix} }

The first building block of our diagrammatic theory is the particle-particle ladder $\varGamma$, which corresponds to the many-body $T$-matrix in the absence of a condensate, and is represented diagrammatically in Figure\til\ref{fig1}. We found it convenient to work with the $T=0$ limit of the Matsubara formalism, that is, we work with imaginary frequencies $i\Omega$.  
 The particle-particle ladder $\varGamma ( \textbf{P}, {\Omega} )$ at momentum $\mathbf{P}$ and imaginary frequency $i{\Omega}$ is given by

\begin{equation} 
\varGamma \left( \textbf{P}, \Omega \right) = \frac{1}{T_2(\textbf{P}, \Omega)^{-1} - I_\mathrm{F}( \textbf{P}, \Omega)},
\end{equation}
where

\begin{align}
 &T_2(\textbf{P},\Omega)^{-1} = -\frac{m_\mathrm{r}}{2\pi}\ln \left( \frac{\textbf{P}^2/(2M) - \mu_\mathrm{F}-\mu_\mathrm{B}-i\Omega}{\epsilon_0}\right) \\
 &I_\mathrm{F}(\textbf{P},\Omega) = \int \frac{\dd^2 k}{(2\pi)^2}\; \frac{\Theta(-\xi_{\textbf{P}-\textbf{k}}^{\mathrm{F}})}{\xi_{\textbf{P}-\textbf{k}}^{\mathrm{F}}+\xi_{\textbf{k}}^{\mathrm{B}}-i\Omega}.
\end{align}
In the above, $T_2(\textbf{P},\Omega)$ is the off-shell two-body $T$-matrix in vacuum and $I_\mathrm{F}(\textbf{P},\Omega)$ stems from the presence of the fermionic component, while $M = m_{\rm B} + m_{\rm F} $. 
We have also assumed $\mu_{\rm B} < 0$, as is the case in the range of couplings, mass ratios, and densities considered.
A closed (albeit lengthy) form for $\varGamma(\textbf{P},\Omega)$ for $\Omega \ne 0$ can be found in \cite{2dbfmix}. Note that for $\mu_{\rm F} < 0$ the second contribution is absent.

Extending the many-body $T$-matrix to the condensed phase requires summing all possible combinations of the repeated scattering between fermions and condensed or non-condensed bosons.

\begin{figure}
    \centering
    \includegraphics[width=1.0\linewidth]{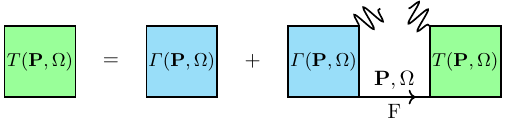}
    \caption{Feynman's diagrams for the many-body $T$-matrix in the condensed phase $T(\mathbf{P},\Omega)$. Wiggly lines correspond to condensed boson insertions and contribute with a factor of $\sqrt{n_0}$ each, where $n_0$ is the condensate density.}
    \label{fig2}
\end{figure}
Thus, the many-body $T$-matrix in the presence of a condensate is shown diagrammatically in Figure\til\ref{fig2}, whose corresponding equation is formally solved to yield
\begin{equation}
    T(\textbf{P},\Omega) = \frac{1}{\varGamma(\textbf{P},\Omega)^{-1}-n_0G_\mathrm{F}^0(\textbf{P},\Omega)}\label{tmatrixdef} ,
\end{equation}
where $n_0$ is the condensate fraction, and $G_s^0(\mathbf{P},\Omega)^{-1} = i\Omega - \xi^s_{\mathbf{P}} $ is the bare Green’s function in the grand-canonical ensemble for the species $s$.

\subsubsection{Boson and fermion self-energies}

\begin{figure}
    \centering
    \includegraphics[width=1\linewidth]{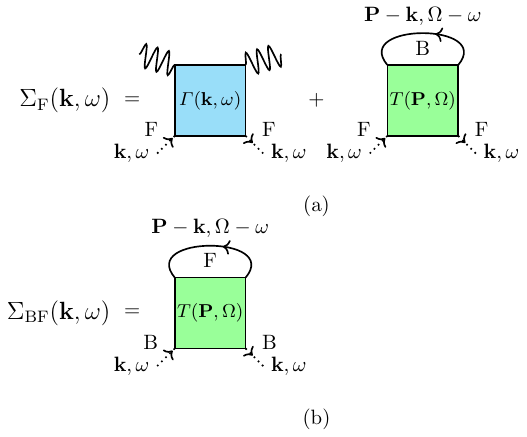}
    \caption{Feynman diagrams for: (a) the fermion self-energy, and (b) the contribution to the normal boson self-energy stemming from BF interactions. The dotted lines with arrows indicate the momentum flow. }
    \label{fig3}
\end{figure}

The fermion self-energy is made of two contributions, represented in Fig.~\ref{fig3}(a),  corresponding to the convolution of the particle-particle ladder $\varGamma(\mathbf{P},\Omega)$ or the 
 many-body $T$-matrix  $T(\mathbf{P},\Omega)$ with a condensate density or a free boson propagator, respectively. The reason for using $\varGamma(\mathbf{P},\Omega)$ rather than $T(\mathbf{P},\Omega)$ in the first diagram of  Fig.~\ref{fig3}(a) is to avoid double counting in the Dyson equation.

One has then
\begin{align} 
\Sigma_\mathrm{F}(\textbf{k},\omega) = n_0 \varGamma(\textbf{k},\omega)&-\int \frac{\dd^2 P}{(2\pi)^2}\int \frac{\dd \Omega}{2\pi}\; T(\textbf{P},\Omega)\nonumber\\
&\times G_\mathrm{B}^0(\textbf{P}-\textbf{k},\Omega-\omega)\e^{i\Omega 0^+}.
 \end{align}

The normal boson self-energy has contributions from both the attractive BF interaction and the (weak) BB repulsion $\Sigma_\mathrm{B}(\textbf{k},\omega) = \Sigma_\mathrm{BF} (\textbf{k},\omega)+\Sigma_{11}$.

The diagram describing the contribution of the BF interaction is shown in Fig.\til\ref{fig3}(b), and corresponds to
\begin{align}
    \Sigma_{\mathrm{BF}}(\textbf{k},\omega) = \int \frac{\dd^2 P}{(2\pi)^2}&\int \frac{\dd \Omega}{2\pi}\; T(\textbf{P},\Omega)\nonumber \\
    \times &G_\mathrm{F}^0(\textbf{P}-\textbf{k},\Omega-\omega)\e^{i\Omega 0^+}.\label{sigmabf}
\end{align}

The inclusion of the BB repulsion gives rise to both normal and anomalous contributions to the bosonic self-energy, which, working at $T = 0$, can be included using the standard Bogoliubov theory (see e.g., \cite{griffin-shi}), or its improved version by Popov \cite{book_popov}. The latter looks more suitable for BF mixtures since a strong BF attraction leads to a large condensate depletion, which is better described by Popov theory. For the same reason, Popov theory was shown in Ref.~\cite{Gualerzi-2025} to better describe the stability properties of a BF mixture in 3D. Thus, we will use the same approach here when the inclusion of the BB interaction is required.

The normal and anomalous contributions to the boson self-energies stemming from the BB interactions are then given respectively by
\begin{align}
    \Sigma_{11} & = \frac{8\pi n_\mathrm{B}\,\eta_\mathrm{B}}{m_\mathrm{B}} \label{normal}\\
    \Sigma_{12}  & = \frac{4\pi n_0\,\eta_\mathrm{B}}{m_\mathrm{B}},\label{anomalous}
\end{align}
where $\eta_{\rm B} = 1/\ln[ 1/(n_{\rm B} a_{\rm BB}^2)]$ is the gas parameter for weakly
interacting bosons in 2D \cite{Schick-1971} and $a_{\rm BB}$ the 2D BB scattering length. 

We notice that most of the numerical results that we will present in the following sections are obtained by neglecting the BB repulsion, unless explicitly stated otherwise. 

\subsubsection{Green's functions, momentum distributions and densities}\label{numerical_methods}

The dressed fermion Green's function $G_{\rm F}(\textbf{k},\omega)$ and the dressed Green's function $G'_{\rm B}(\textbf{k},\omega)$ for non-condensed bosons are obtained by inserting the (irreducible) diagrams shown in Fig.\til\ref{fig3} into Dyson's equations
\begin{align}
    G_\mathrm{F}(\textbf{k},\omega)^{-1} &= G_\mathrm{F}^0(\textbf{k},\omega)^{-1}-\Sigma_\mathrm{F}(\textbf{k},\omega)\\
    G'_\mathrm{B}(\textbf{k},\omega)^{-1} &= G_\mathrm{B}^0(\textbf{k},\omega)^{-1}-\Sigma_\mathrm{B}(\textbf{k},\omega)\nonumber\\
    &\hspace{0.5cm}+\frac{\Sigma_{12}^2}{i\omega+\xi_\textbf{k}^\mathrm{B}+\Sigma_\mathrm{B}(-\textbf{k},-\omega)} .
\end{align}

The momentum distributions for fermions and non-condensed bosons are then given by
\begin{align}
     &n_\mathrm{F}(\textbf{k}) = \int \frac{\dd \omega}{2\pi}\, G_\mathrm{F}(\textbf{k},\omega)\e^{i\omega 0^+}\\
    &n_\mathrm{B}(\textbf{k}) =-\int \frac{\dd \omega}{2\pi}\, G'_\mathrm{B}(\textbf{k},\omega)\e^{i\omega 0^+}.
\end{align}

\subsection{Numerical methods}\label{numimplement}
The core of our numerical calculations is constituted by the set of equations
\begin{align}
    &n_\mathrm{F} = \int \frac{\dd^2 k}{(2\pi)^2} \,n_\mathrm{F}(\textbf{k})\label{nf}\\
    &n_\mathrm{B} =n_0 + \int \frac{\dd^2 k}{(2\pi)^2}\, n_\mathrm{B}(\textbf{k})\label{nb}\\
    &\mu_\mathrm{B} = \Sigma_\mathrm{B}(\textbf{0},0)\label{hp}-\Sigma_{12}.
\end{align}
Equations \til\eqref{nf}-\eqref{nb} are the standard equations relating the momentum distribution to the number density of a many-body assembly, while Eq.~\eqref{hp} is the Hugenholtz-Pines condition for a system with a boson condensate \cite{HPeq,Hohenberg-1965,Rickayzen-1980}. For given set of parameters $x$, $\gamma_\mathrm{m}$, $\eta_\mathrm{B}$ and $g$, the above system is solved for the unknowns $\mu_\mathrm{F}$, $\mu_\mathrm{B}$ (both in units of $E_\mathrm{F}$) and $n_0$ (in units of $n_\mathrm{F}$).

As discussed in \cite{2dbfmix} for equal masses, in the strong-coupling limit $g\gtrsim 2$, the system of Eqs.~\eqref{nf}-\eqref{hp} can be replaced by the simpler set of equations
\begin{align}
    &  n_\mathrm{F} = \int\frac{\dd^2 k}{\left( 2\pi \right)^2} \;n_\mathrm{F} ( \textbf{k} ) \label{nfsc}\\
    & n_\mathrm{B}-n_0= \int\frac{\dd^2 P}{(2\pi)^2}\; \left( u_{\textbf{P}}^2 \Theta(-E_\textbf{P}^+) +  v_\textbf{P}^2 \Theta(-E_\textbf{P}^-)  \right) \label{nbsc}\\
    &\mu_\mathrm{B}= \frac{2 \pi \epsilon_0}{m_\mathrm{r}} \int \frac{\dd^2 P}{(2\pi)^2} \bigg(  \frac{u_\textbf{P}^2}{E_\textbf{P}^+ - \xi^\mathrm{F}_{\textbf{P}}} \Theta(-E_\textbf{P}^+)\nonumber\\ &\hspace{2.9cm}+  
           \frac{ v_\textbf{P}^2 }{  E_\textbf{P}^- - \xi^\mathrm{F}_{\textbf{P}}} \Theta(-E_\textbf{P}^-)     \bigg) \label{hpsc},
\end{align}
which is again solved for $\mu_{\rm F}, \mu_{\rm B}$ and $n_0$. 
In the above equations, $E_\textbf{P}^\pm$ denote the dispersion relations of the poles of the many-body $T$-matrix in the strong-coupling limit, and $u_\textbf{P}^2$, $v_\textbf{P}^2$ their corresponding quasi-particle weights. Explicit expressions for $E_\textbf{P}^\pm$ and $u_\textbf{P}^2$, $v_\textbf{P}^2$ can be found in Sec.~\ref{hybsec}, while the closed-form solutions for the integrals in Eqs.~\eqref{nbsc} and \eqref{hpsc} can be found in Appendix~\ref{asssc}. A strong-coupling expression for the fermion momentum distribution in Eq.~\eqref{nfsc} is also reported in Appendix~\ref{asssc}.

\section{Numerical results for \texorpdfstring{$n_\mathrm{B}=n_\mathrm{F}$}{nB = nF}}\label{sec:xuno}

\subsection{Chemical potentials}\label{sec:numchempot}

\begin{figure}[ht]
    \centering
    \includegraphics[width=1.0\linewidth]{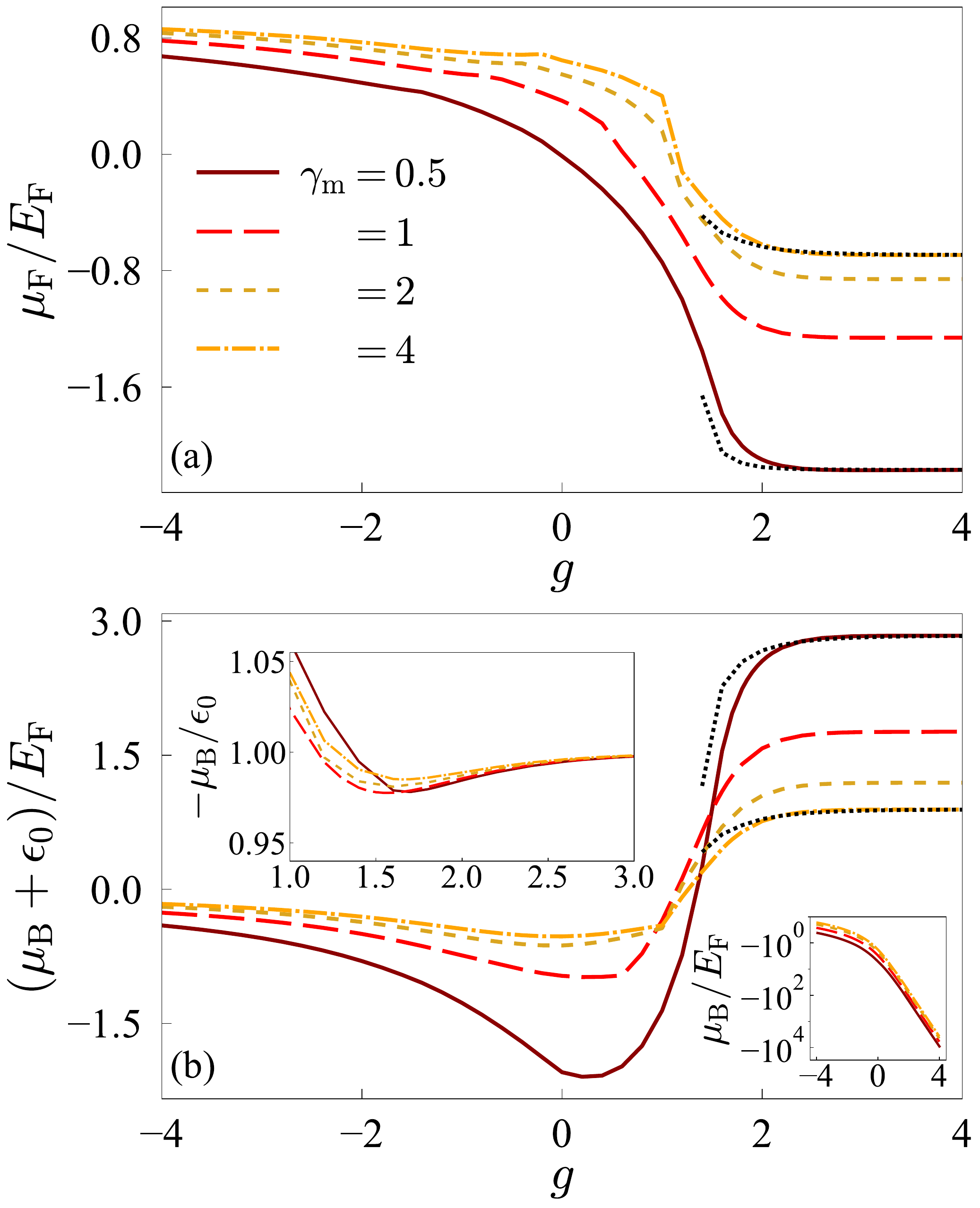}
         \caption{(a):  Fermion chemical potential $\mu_\mathrm{F}$ in units of $E_\mathrm{F}$, as a function of the BF coupling $g$, for several values of mass ratio $\gamma_\mathrm{m}$, in the case $x = 1$. (b): Corresponding boson chemical potential $\mu_\mathrm{B}$ (shifted by the binding energy $\epsilon_0$), in units of $E_\mathrm{F}$. Top inset:  ratio between $\mu_\mathrm{B}$ and its leading strong-coupling limit  $-\epsilon_0$. Black dotted lines: strong-coupling benchmarks (reported for clarity for $\gamma_\mathrm{m} = 0.5, 4$ only). Bottom inset: bosonic chemical potential in units of $E_{\rm F}$ as a function of $g$. 
         }
    \label{fig4}
\end{figure}

Figure\til\ref{fig4} presents the evolution of the mixture's chemical potentials from weak to strong coupling ($-4\le g \le 4$) for the selected case $x= 1$.  Note that for the boson chemical potential, we report in the main plot the quantity $\mu_\mathrm{B}+\epsilon_0$, to subtract the leading strong-coupling contribution given by $-\epsilon_0/E_\mathrm{F} = - \e^{2g} (\gamma_\mathrm{m}+1)/\gamma_\mathrm{m}$.

One can see that changing the mass ratio significantly modifies the quantitative values of the chemical potentials without altering, however, their qualitative behavior. 
In particular, both the fermionic and bosonic chemical potentials monotonically decrease with increasing interaction strength, as shown in Fig.\til\ref{fig4}(a) and the bottom inset in Fig.\til\ref{fig4}(b), respectively. 
We have observed a qualitatively similar behavior also for $x < 1$: varying the mass ratio changes the chemical potentials only quantitatively with respect to the corresponding quantities for $\gamma_\mathrm{m} = 1$ reported in \cite{2dbfmix}.

In the top inset of panel (b), we report the ratio between $\mu_{\rm B}$ and its leading strong-coupling limit $-\epsilon_0$. One sees that the relative difference $|\mu_\mathrm{B}+\epsilon_0|/\epsilon_0$ goes below $5\%$ for $g \simeq 1$ for all the mass ratios considered. 

The numerical results for the two chemical potentials are also compared with the solution of the asymptotic strong-coupling equations \eqref{nfsc}-\eqref{hpsc}.
The good agreement with the strong-coupling benchmarks (black dotted lines) demonstrates that the assumptions underlying the strong-coupling theory developed in \cite{2dbfmix} for $\gamma_\mathrm{m}=1$ remain valid also for $\gamma_\mathrm{m}\neq1$. 

In the opposite weak-coupling limit $1/g\to 0^-$, the controlled perturbative expansion of Ref.~\cite{montecarlo} provides a benchmark for the values of the chemical potentials.
In Appendix~\ref{appwc} we compare the above second-order expansion with the numerical results obtained within the present $T$-matrix formalism, thus verifying that in the weak-coupling limit our theoretical approach is correct to second order in $1/g$.
In addition, we provide a detailed expansion of our set of equations in the weak-coupling limit, thus showing analytically how the perturbative results for the chemical potentials are recovered by the present approach.

\begin{figure}[t]
    \centering
    \includegraphics[width=1.0\linewidth]{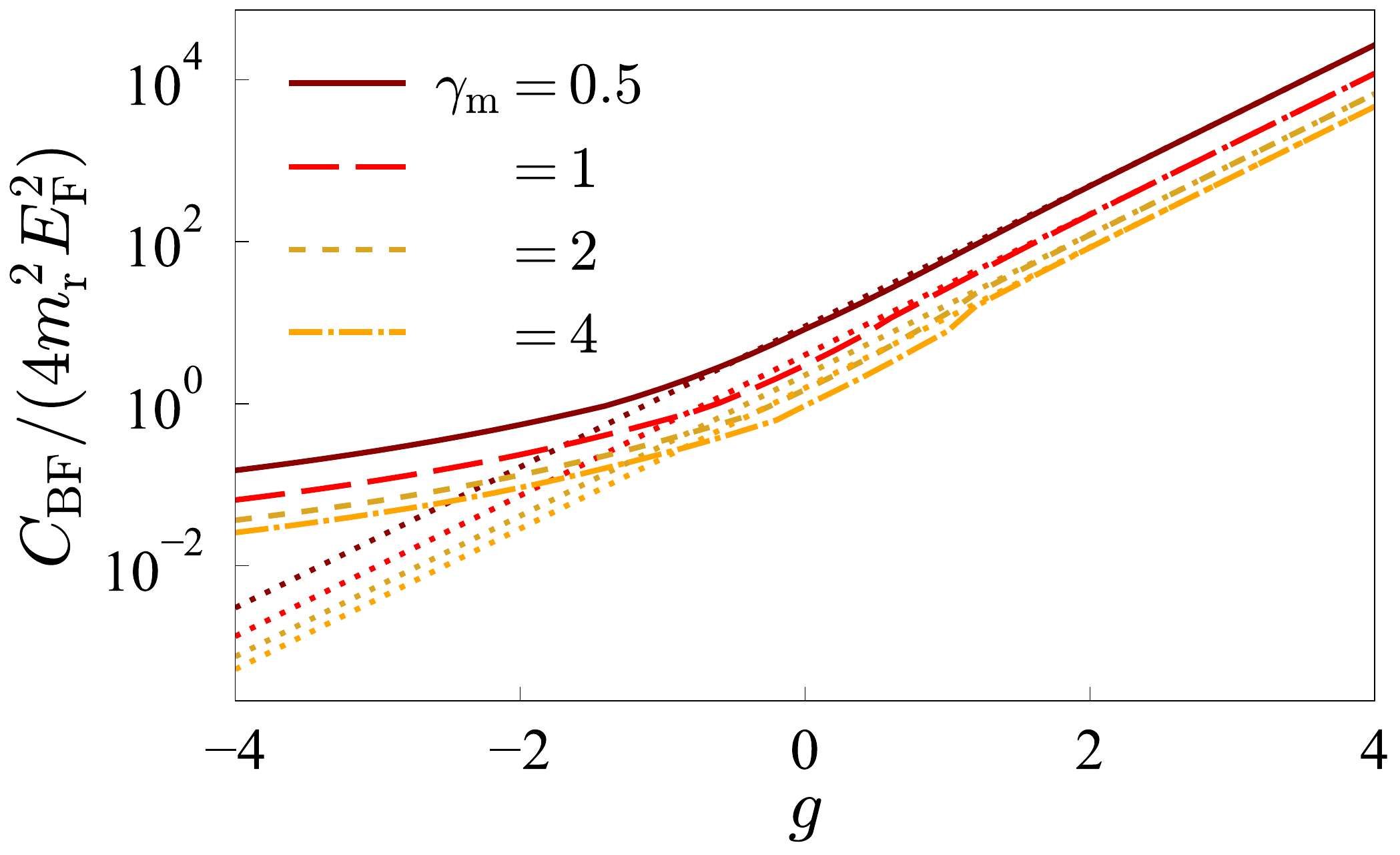}
         
    \caption{Dimensionless parameter $C_\mathrm{BF}/(4m_\mathrm{r}^2E_\mathrm{F}^2)$, as a function of BF coupling $g$, for different values of the mass ratio $\gamma_\mathrm{m}$ in the case $x = 1$. Dotted lines: asymptotic behavior given by $\left(\epsilon_0/E_\mathrm{F}\right) \, (\gamma_\mathrm{m}+1)/\gamma_\mathrm{m}$.}
    \label{fig5}
\end{figure}

\subsection{Tan's contact parameter}\label{subsec:tan}
Quite generally, for contact interactions, the fermion and boson momentum distributions at large momenta are given by  $n_s(\textbf{k}) = C_s/k^4$ to leading order (where $k=|{\bf k}|$) \cite{Tan-2008a,Tan-2008b,Tan-2008c}. Here, $C_s$ is the so-called Tan's contact parameter for the species $s$.
Within our theory, in the absence of BB repulsion, one has $C_{\rm B}=C_{\rm F}=C_{\rm BF}$, where we have introduced 
the quantity
\begin{equation}
C_\mathrm{BF} = {4m_\mathrm{r}^2}\int \frac{\dd^2 P}{(2\pi)^2}\int \frac{\dd \Omega}{2\pi}\,T(\textbf{P},\Omega)\e^{i\Omega 0^+}
\end{equation}
that has the meaning of the contribution to Tan's contact originating from the BF interaction.
In the presence of a BB repulsion, the boson distribution acquires an additional contribution originating from the anomalous self-energy, yielding $C_{\rm B}=C_\mathrm{BF} + \left(m_\mathrm{B}\Sigma_{12}\right)^2$, while $C_{\rm F}$ remains equal to $C_{\rm BF}$.

The dimensionless quantity  $C_\mathrm{BF}/(4m_\mathrm{r}^2E_\mathrm{F}^2)$ for $x = 1$ is shown in Fig.~\ref{fig5}, from weak to strong coupling and for several values of the mass ratio. For $\gamma_{\rm m}=4$, one sees in Fig.~\ref{fig5} a rapid change in the slope of the curve near $g \simeq 1.2$.  Although less visible, the same change in behavior also appears for the remaining mass ratios. For all cases, we have verified that this feature in the Tan's contact occurs at the coupling where $\mu_{\rm F}$ becomes negative (see Fig.\til\ref{fig4}(a)).  This change of sign can be interpreted as marking a crossover from a regime with a Fermi surface of predominantly unpaired fermions to a molecular regime increasingly dominated by bound BF pairs, in which Tan’s contact approaches its strong-coupling behavior.

In the strong-coupling limit, one can obtain the following asymptotic behavior for $x=1$:
\begin{equation}
        \frac{C_\mathrm{BF}}{4 m_\mathrm{r}^2 E_\mathrm{F}^2} \sim \frac{\gamma_\mathrm{m}+1}{\gamma_\mathrm{m}}\frac{\epsilon_0}{E_\mathrm{F}}= \left(\frac{\gamma_\mathrm{m}+1}{\gamma_\mathrm{m}}\right)^2 \e^{2g}.\label{trace}
\end{equation}

The dotted lines in Fig.~\ref{fig5} represent the asymptotic prediction \eqref{trace}, showing good agreement with the numerical results in the strong-coupling regime for all mass ratios considered.

We have observed similar good agreement also for $x<1$. In this case, the asymptotic expression on the right-hand side of Eq.~\eqref{trace} is multiplied by  $x$. This prefactor follows from the fact that, in the strong-coupling limit, the $T$-matrix reduces to the molecular propagator (up to a constant) \cite{2dbfmix}. Thus, the integral of the $T$-matrix over momentum and frequency is proportional to the density of composite fermions. Since in the strong coupling limit the condensate is mostly depleted, one can approximate the density of BF pairs with the boson density to leading order, obtaining the prefactor $x$.

\begin{figure}[t]
    \centering
    \includegraphics[width=1.0\linewidth]{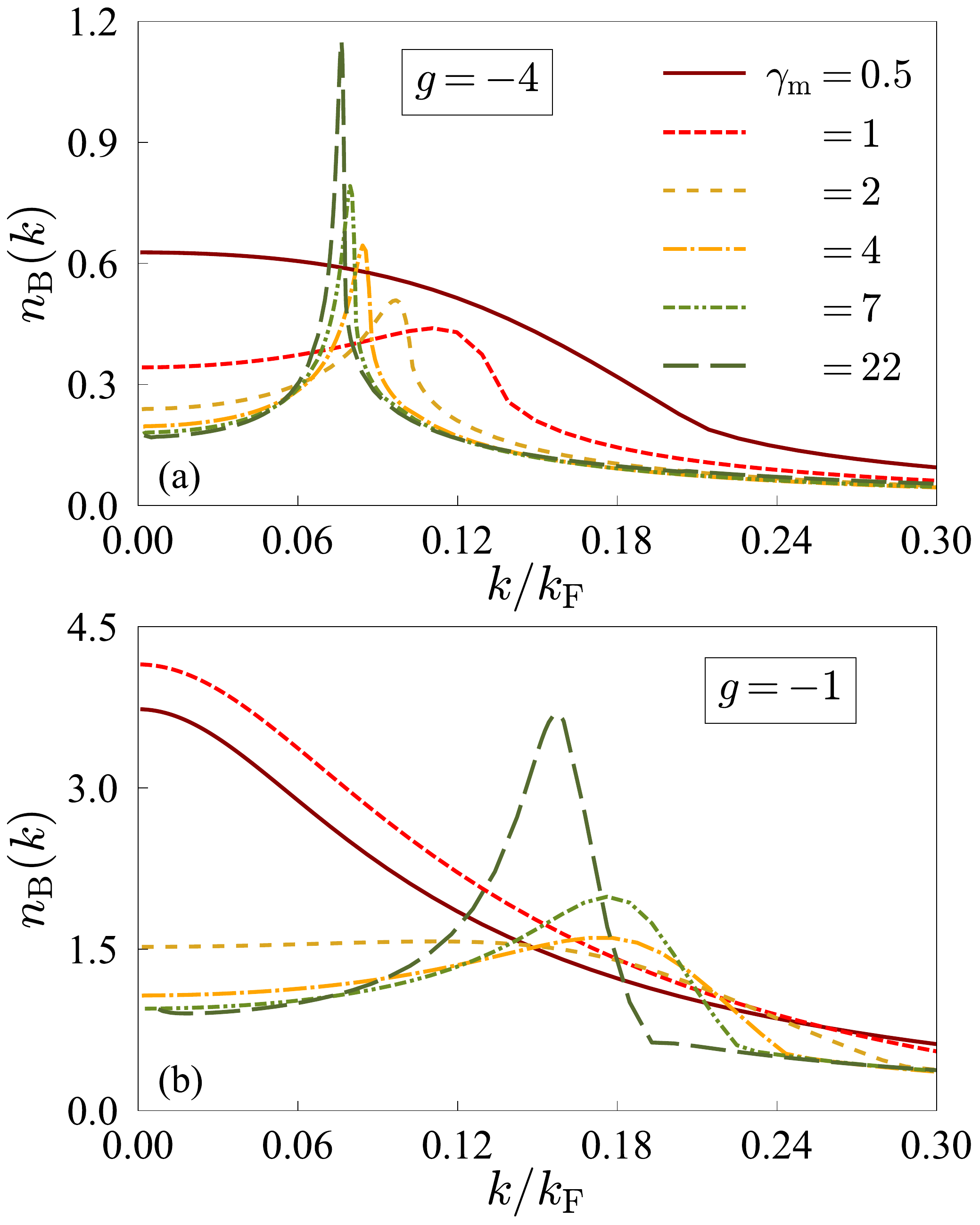}
         \caption{Boson momentum distribution $n_\mathrm{B}(k)$ as a function of $k/k_\mathrm{F}$, for the boson concentration $x = 1$ and BF couplings $g = -4$ (a) and $g = -1$ (b). Several values of 
    mass ratio $\gamma_\mathrm{m}$ are reported.}
    \label{fig6}
\end{figure}

\subsection{Boson momentum distribution}

An interesting feature of the bosonic momentum distribution for equal masses obtained in \cite{2dbfmix}, is the presence of a peak at finite momentum, which is observed only in the weak-coupling regime. In particular, for the density-matched mixture ($x=1$), the peak is present for $g = -4$, while it is absent already for $g = -2$.
We now explore the effect of the bosonic mass on this feature.

Fig.~\ref{fig6} shows the boson momentum distribution for $x = 1$ for several values of mass ratio and BF coupling.

In the upper panel, corresponding to the weak-coupling case $g = -4$, the boson momentum distribution exhibits a maximum at non-zero momentum for all cases in which the boson mass is equal to or larger than the fermion one (corresponding to $\gamma_{\rm m} \ge 1$). 
In particular, the peak shifts to lower momentum while becoming more prominent as the mass ratio increases.

By contrast, the lower panel shows that, for sufficiently large values of $\gamma_\mathrm{m} \ge 4 $, a well-defined finite-momentum peak is present even outside the weak-coupling regime ($g = -1$).
Increasing the BF coupling shifts the peak to larger momenta and increases its height.

An ongoing analysis of the spectral-weight functions \cite{Bovini-2026} indicates that the finite-momentum peak originates from an avoided crossing between two pole branches of the bosonic Green’s function. This avoided crossing produces a nonmonotonic momentum dependence of the spectral weight carried by the negative-frequency branch, which provides the dominant contribution to $n_{\rm B}(k)$ at small momenta. Specifically, this weight first increases and then rapidly vanishes as the negative-frequency branch crosses zero frequency, giving rise to the peak followed by a kink. The kink occurs at the momentum where the branch crosses zero frequency and is most clearly visible for $\gamma_{\rm m}=4$, 7, and 22 in Fig.~\ref{fig6}(b).
As the coupling increases, the two pole branches move progressively farther apart in energy, so that the avoided crossing and the finite-momentum peak eventually disappear at sufficiently strong coupling. The zero-frequency crossing, and hence the associated kink, generally persists, although the kink is not always clearly visible.

\section{Numerical results for   \texorpdfstring{$n_\mathrm{B}\le n_\mathrm{F}$}{ nB less nF}  }\label{ximnoreuno}

\subsection{Boson momentum distribution}

\begin{figure}[ht]
    \centering
    \includegraphics[width=1.0\linewidth]{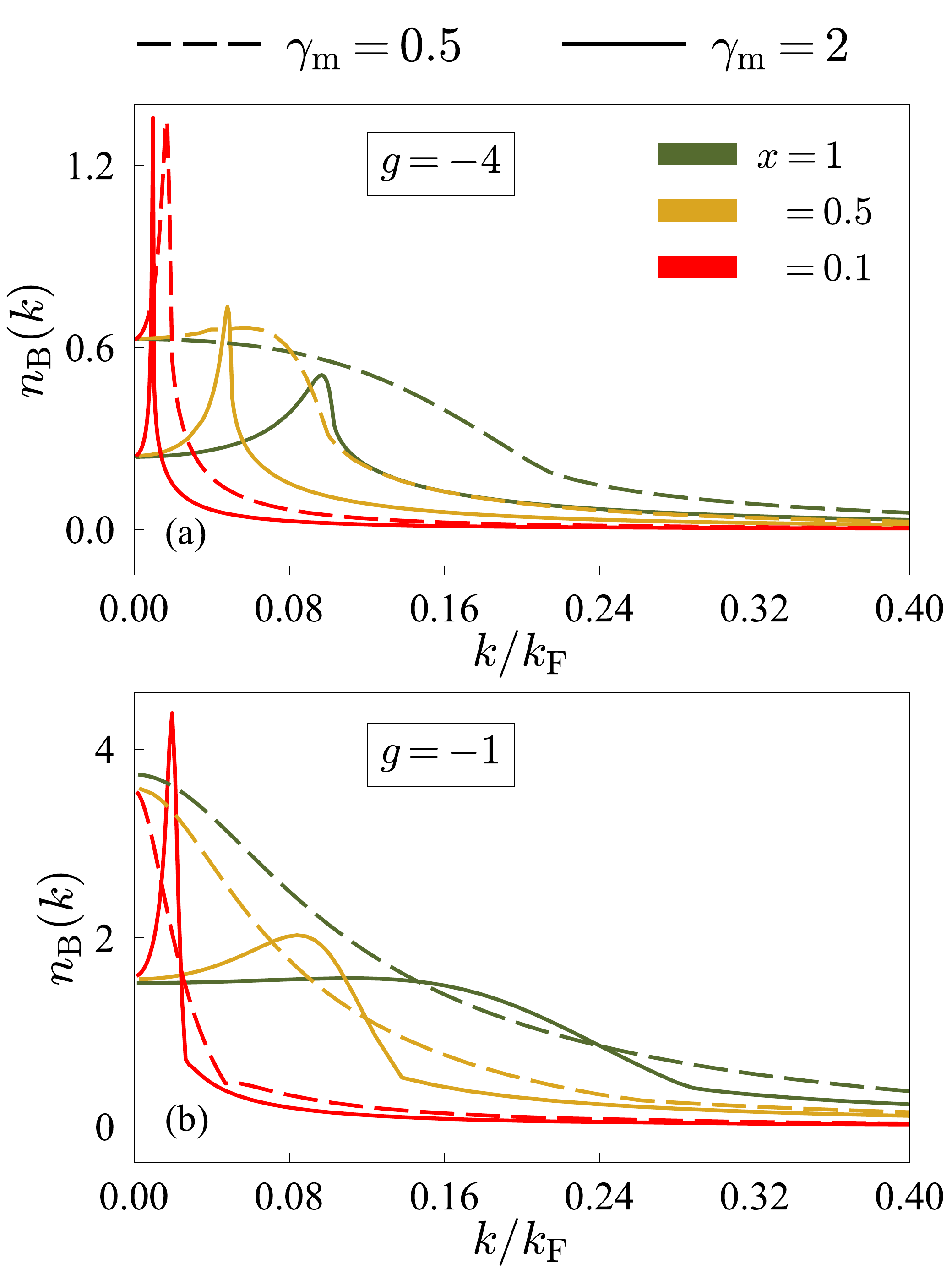}
         
    \caption{Boson momentum distribution $n_\mathrm{B}(k)$ as a function of $k/k_\mathrm{F}$, for BF couplings $g = -4$ (a) and $g = -1$ (b). Several different values of the boson concentration $x$ are reported: $x = 1$ (green), $x = 0.5$ (yellow) and $x = 0.1$ (red). Two different values of the mass ratio $\gamma_\mathrm{m}$ are considered: $\gamma_\mathrm{m} = 0.5$ (dashed lines) and $\gamma_\mathrm{m} = 2$ (full lines).}
    \label{fig7}
\end{figure}

Figure\til\ref{fig7} illustrates the dependence of the bosonic momentum distribution $n_\mathrm{B}(\mathbf{k})$ on the boson concentration $x$ in the low-momentum region near the peak (when present) discussed in the previous section. We report the same two BF coupling values considered in Fig.~\ref{fig6}, for two selected mass ratios.

The results of Fig.~\ref{fig7} for $x < 1$ together with those of Fig.~\ref{fig6} at $x=1$ reveal a general interplay between BF coupling, boson concentration, and mass ratio. Increasing the BF coupling tends to broaden the finite-momentum peak (when present) and shift it toward larger momenta. Conversely, increasing the mass ratio $\gamma_{\rm m}$ leads to a sharper peak located at lower momentum. Finally, reducing the boson concentration generally enhances the peak's visibility and further shifts it toward smaller momenta. 

Measurements of the bosonic momentum distribution in two-dimensional BF mixtures in future experiments could provide a direct test of our predictions. In ultracold-atom realizations, such measurements could be performed through time-of-flight imaging.

However, a remark concerning a possible experimental test of our predictions is in order.
All of the above results neglect the presence of a BB repulsion. It is well known that, in the absence of a BB repulsion, a homogeneous BF mixture is mechanically unstable toward collapse or phase separation \cite{Viverit-2000,Zhai-2011,Gualerzi-2025,Cordioli-2026}. As such, our results with BB repulsion $\eta_{\rm B}=0$ could be applied only to experiments fast enough compared with the timescales over which the instability sets in. This is what, according to its authors, occurred in the experiment \cite{Bloch-2023}, which successfully explored a 3D Bose-Fermi mixture with a negligible BB repulsion across the whole evolution of the BF coupling from weak to strong.

Still, it would be desirable to work with, and possibly observe the above peak in, mechanically stable mixtures. In \cite{2dbfmix} it was observed that while the inclusion of a repulsion $\eta_{\rm B}$ does not qualitatively affect the thermodynamic quantities $\mu_{\rm B}, \mu_{\rm F}, n_0$, it may have a large effect on the momentum distribution of bosons at small momenta. 
In particular, its inclusion causes the momentum distribution to diverge as $n_{\rm B}(\textbf{k}) \sim 1/k$ for $k \to 0$. 
The behavior of the momentum distribution in the presence of Bose-Bose repulsion can be understood as follows. In our approach, the Bose-Bose repulsion is incorporated through the normal and anomalous self-energies $\Sigma_{11}$ and $\Sigma_{12}$ given by Eqs.~\eqref{normal} and \eqref{anomalous}. When the Bose-Fermi interaction is switched off, the bosonic normal Green's function reduces to the standard Bogoliubov form, $G'_{\rm B}(k,\omega)=u_k^2/(i\omega-E_k)-v_k^2/(i\omega+E_k)$, which yields $n_{\rm B}(k)=v_k^2\sim 1/k$ as $k\to0$. This low-momentum divergence is expected to persist when the Bose-Fermi contribution $\Sigma_{\rm BF}$ is included in $\Sigma_{11}$, as confirmed by our numerical calculations.

This divergence could completely mask the finite-momentum peak discussed so far (similarly to what was observed for equal masses in Fig.\til10 of \cite{2dbfmix}).

\begin{figure}[t]
    \centering
    \includegraphics[width=1.0\linewidth]{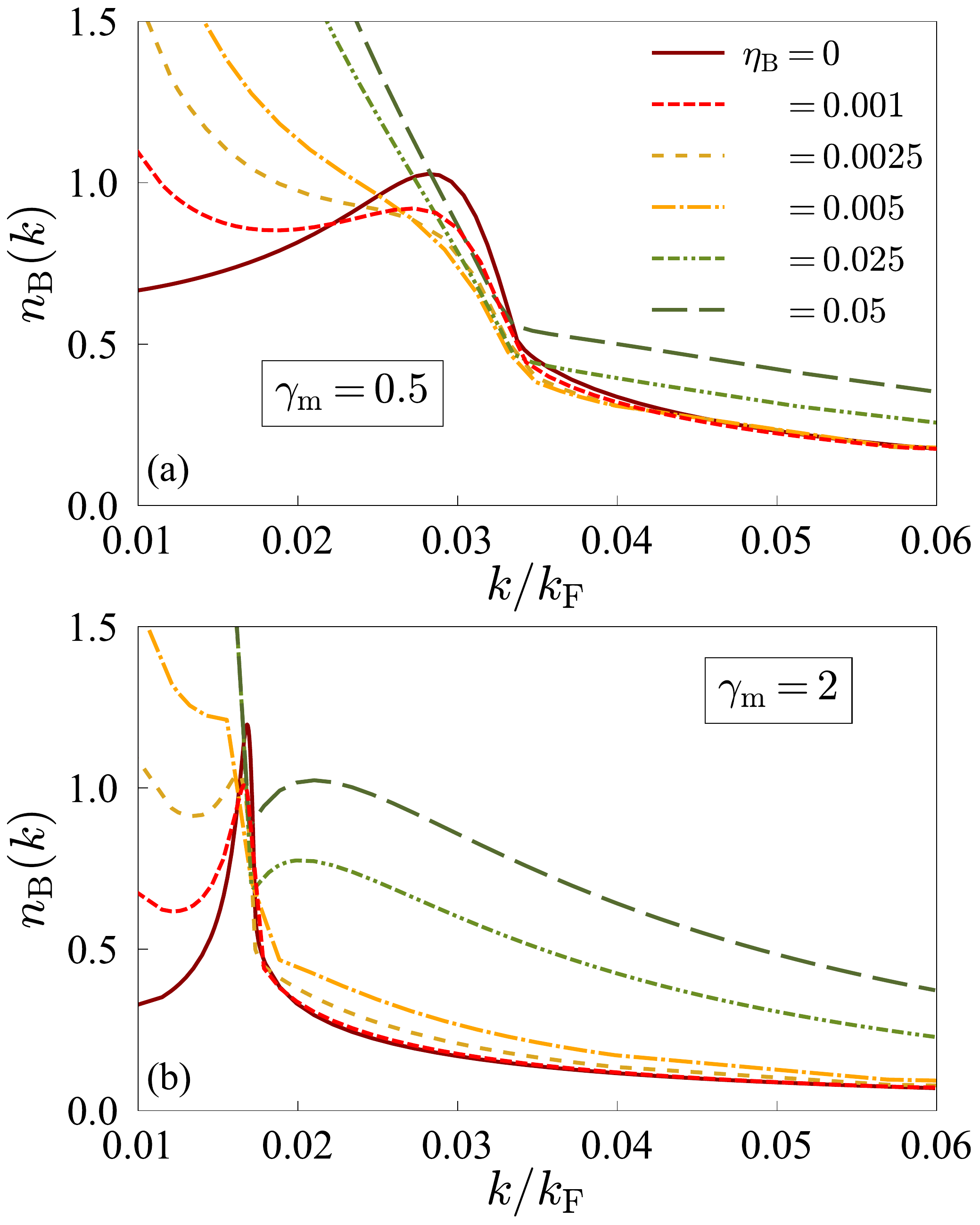}
         
    \caption{Boson momentum distribution $n_\mathrm{B}(k)$ as a function of $k/k_\mathrm{F}$, for the BF coupling $g = -4$, boson concentration $x = 0.175$ and several different values of the BB coupling $\eta_\mathrm{B}$. Two values for the mass ratio $\gamma_\mathrm{m}$ are reported: $\gamma_\mathrm{m} = 0.5$ (a) and $\gamma_\mathrm{m} = 2$ (b). }
    \label{fig8}
\end{figure}

In Fig.~\ref{fig8} we show the boson momentum distribution in a selected case $x=0.175$ in the weak-coupling regime $g = -4$ for the same two values of $\gamma_{\rm m}$ as in Fig.~\ref{fig7}, with the BB repulsion parameter varying from $\eta_{\rm B} = 0$ to $\eta_{\rm B} = 0.05$. This value should correspond to a BB repulsion strong enough to render the mixture mechanically stable according to the analysis of \cite{Cordioli-2026}.

One sees that in the case of a smaller boson mass ($\gamma_{\rm m}=0.5$), the introduction of a finite BB repulsion rapidly washes out the finite-momentum peak.
For the larger value $\gamma_{\rm m}=2$, a peak at finite momentum is instead clearly visible even for the largest values of the BB repulsion ($\eta_{\rm B}=0.025$ and $\eta_{\rm B}=0.05$), for which the mixture should be mechanically stable \cite{Cordioli-2026}.  More generally, mixtures with mass ratios $\gamma_{\rm m} \gtrsim 2$, BF coupling $-4 \lesssim g \lesssim -2$ and BB repulsion $0.025 \lesssim \eta_{\rm B} \lesssim 0.05$ should allow for the observability of the predicted finite momentum peak in a mechanically stable BF mixture.

\begin{figure}[t]
    \centering
    \includegraphics[width=1.0\linewidth]{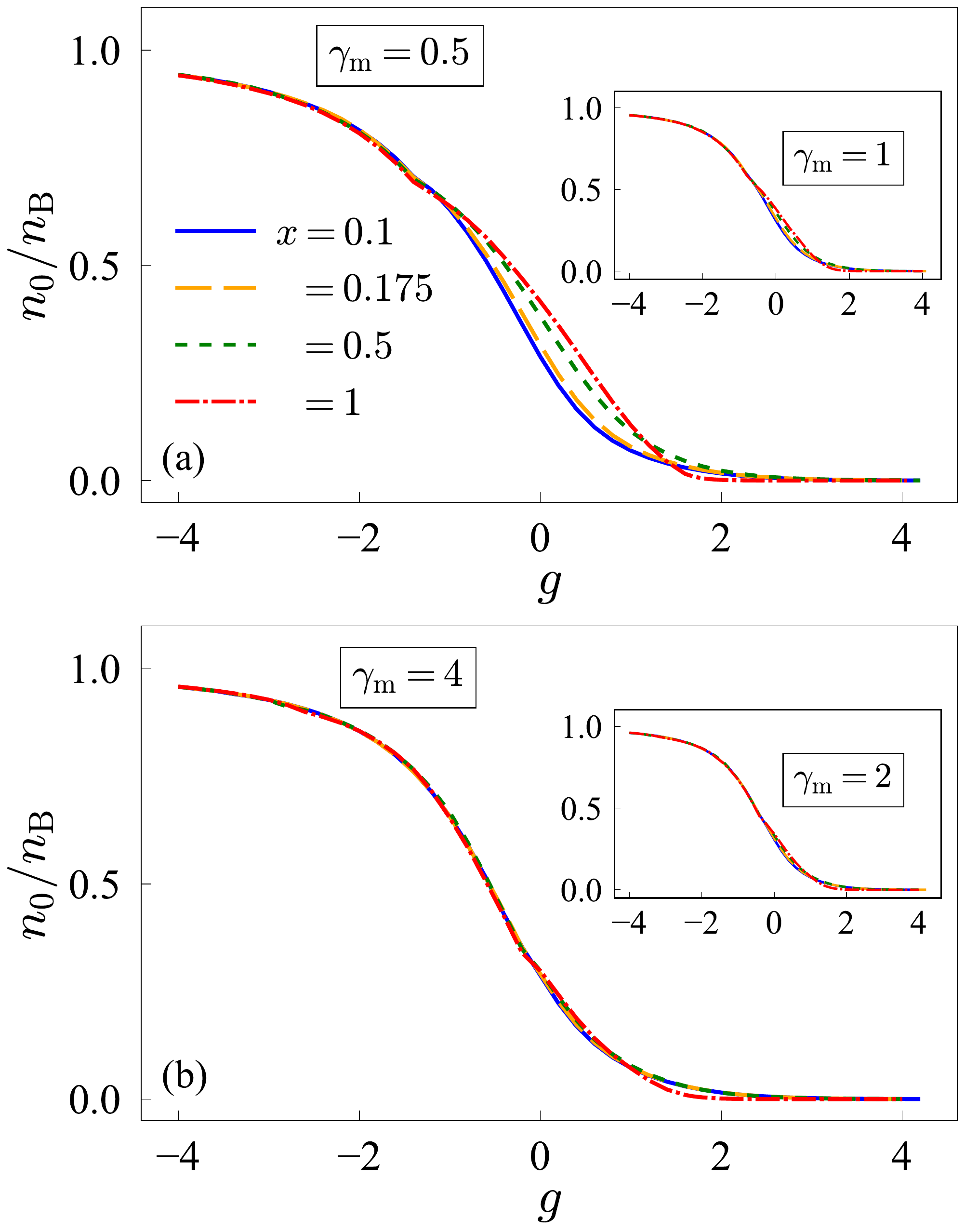}
         
    \caption{Boson condensate fraction $n_0/n_\mathrm{B}$ as a function of the BF coupling parameter $g$ for different values of the boson concentration $x$ and mass ratio $\gamma_\mathrm{m}$: (a) $\gamma_\mathrm{m} = 0.5$ (inset $\gamma_\mathrm{m} = 1$), (b) $\gamma_\mathrm{m} = 4$ (inset $\gamma_\mathrm{m} = 2$).}
    \label{fig9}
\end{figure}

\subsection{Universality of the boson condensate fraction}

A near-universal dependence of the condensate fraction on the boson density was observed in Ref.~\cite{2dbfmix} for equal masses, similarly to what was previously found in \cite{Guidini-2015} in three dimensions.
Here, we investigate the effects of a mass imbalance.

Figure\til\ref{fig9} shows the condensate fraction as a function of the BF coupling $g$ for several values of boson concentrations and mass ratios.
The overall behavior is qualitatively similar to that obtained for equal masses \cite{2dbfmix}, with a monotonic decrease of the condensate fraction as the BF attraction increases, the absence of a critical coupling past which the condensate vanishes identically, and a near-universal dependence on the boson density. Quantitative deviations are, however, present. In particular, deviations from universality are more pronounced for $\gamma_{\rm m}<1$ (upper panel), whereas they are reduced for $\gamma_{\rm m}>1$ (lower panel and inset), similarly to what was previously observed in the three-dimensional case \cite{Guidini-2015}.

\begin{figure}
    \centering
    \includegraphics[width=1\linewidth]{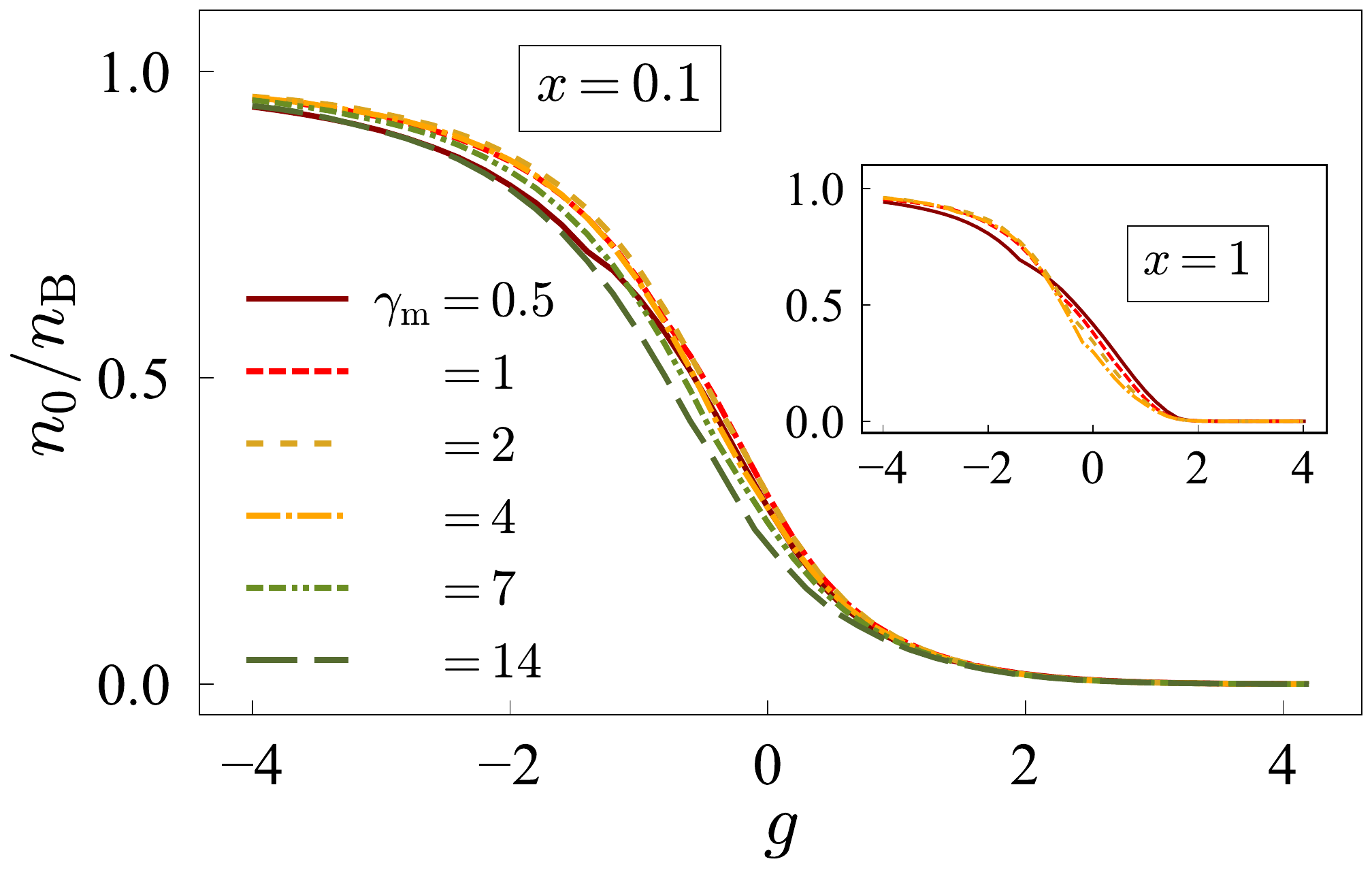}
    \caption{Boson condensate fraction $n_0/n_\mathrm{B}$ as a function of the BF coupling parameter $g$ for several values of mass ratio $\gamma_\mathrm{m}$ and for boson concentration $x = 0.1$ (main panel) and $x = 1$ (inset).}
    \label{fig10}
\end{figure}

To further highlight the effects of changing the mass ratio, Fig.~\ref{fig10} reports $n_0/n_\mathrm{B}$ as a function of the BF coupling $g$ for several values of the mass ratio at two different concentrations: $x=0.1$ (main panel) and $x=1$ (inset).
One sees in Fig.~\ref{fig10} a non-trivial dependence of the condensate fraction on the mass ratio for both concentrations. For intermediate to strong interactions, the condensate is more depleted at large boson masses. In the weak-coupling region, the dependence on the boson mass is instead non-monotonic.  
In particular, we have numerically verified the following second-order perturbative expansion for the boson condensate density \cite{Cardarelli-2014}
\begin{equation}
n_0 = n_\mathrm{B}\left[1 - \frac{1}{4g^2}\left(\frac{\gamma_\mathrm{m} + 1}{\gamma_\mathrm{m}}\right)^2 \ln(\gamma_\mathrm{m}+1)\right],
    \label{n0pert}
\end{equation}
from which one immediately sees that the condensate depletion increases as $1/\gamma_{\rm m}$ for small $\gamma_{\rm m}$ and as $\ln \gamma_{\rm m}$ at large $\gamma_{\rm m}$, with a minimum for $\gamma_{\rm m} \simeq 2.5$.

\subsection{Hybridization energy and
 poles of the many-body \texorpdfstring{$T$-matrix}{T matrix} in the strong-coupling limit}\label{hybsec}

In the strong-coupling limit, the many-body $T$-matrix acquires the meaning of a composite fermion propagator~\cite{2dbfmix}. In particular, in the condensed phase, it acquires the following two-pole structure 
\begin{align}
    &T(\mathbf{P},\Omega) =\frac{2\pi\epsilon_0}{m_{\rm r}}\left(\frac{u_\textbf{P}^2}{i\Omega - E_\textbf{P}^+}+ \frac{v_\textbf{P}^2}{i\Omega - E_\textbf{P}^-}\right),
\end{align} 
with dispersions 
\begin{align}
    & E_\textbf{P}^\pm = \frac{\widetilde{\xi}_\textbf{P}^{\mathrm{CF}}+\xi_\textbf{P}^\mathrm{F}\pm \sqrt{\left(\widetilde{\xi}_\textbf{P}^{\mathrm{CF}}-\xi_\textbf{P}^\mathrm{F}\right)^2+4\Delta_0^2}}{2} 
    \label{hybrid}
\end{align}
and weights
\begin{align}
    &u_\textbf{P}^2 = \frac{1}{2}\left(1+\frac{\widetilde{\xi}_\textbf{P}^{\mathrm{CF}}-\xi_\textbf{P}^\mathrm{F}}{\sqrt{\left(\widetilde{\xi}_\textbf{P}^{\mathrm{CF}}-\xi_\textbf{P}^\mathrm{F}\right)^2 + 4 \Delta_0^2}}\right),\\
    &v_\textbf{P}^2 = 1 -u_\textbf{P}^2 .
\end{align} 

One sees in Eq.~(\ref{hybrid}) that the energy scale $\Delta_0\equiv(2\pi \epsilon_0 n_0/m_{\rm r})^{1/2}$
controls the amount of hybridization between a free fermion dispersion $\xi_\textbf{P}^\mathrm{F}$  
and a molecular dispersion
$\widetilde{\xi}_\textbf{P}^{\,\mathrm{CF}} = \textbf{P}^2/(2M)-\widetilde{\mu}_\mathrm{CF}$  with chemical potential shifted by a mean-field contribution stemming from the presence of a degenerate Fermi component: $\widetilde{\mu}_\mathrm{CF} = \mu_\mathrm{B}+\mu_\mathrm{F}+\epsilon_0 - \Sigma_\mathrm{CF}^0$ with $\Sigma_\mathrm{CF}^0 =   m_\mathrm{F} \mu_\mathrm{F} \Theta(\mu_\mathrm{F})/m_\mathrm{r}$.

\begin{figure}[t]
    \centering
    \includegraphics[width=1.0\linewidth]{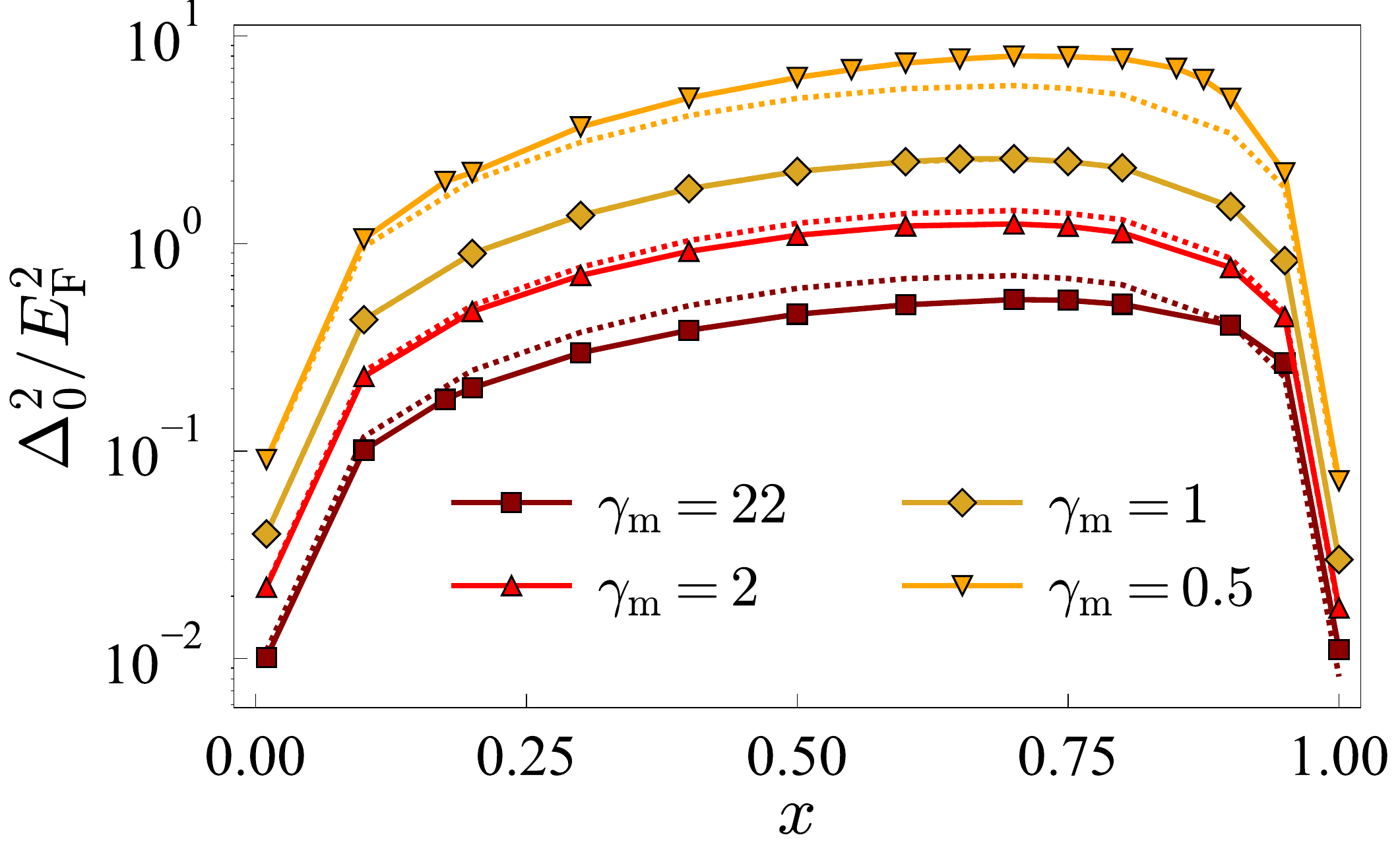}
         
    \caption{Square of the hybridization energy $\Delta_0^2 =2\pi \epsilon_0 n_0/m_{\rm r}$ (in units of $E_\mathrm{F}^2$) as a function of the boson concentration $x$, for strong BF attraction $g = 3$ and different mass ratios. Dotted lines: same quantity with $n_0$ kept at its value for $\gamma_{\rm m} =1$.}
    \label{fig11}
\end{figure}

Figure\til\ref{fig11} illustrates the effects of mass imbalance on the hybridization energy scale $\Delta_0$ for a fixed strong BF coupling and several values of the boson concentration. One sees that $\Delta_0/E_{\rm F}$ decreases monotonically as the mass ratio increases for all values of $x$ considered. 
This behavior is easily understood by noting that
$\Delta_0/E_{\rm F}=\e^g\sqrt{ x \, n_0/n_{\rm B}}(1+1/\gamma_{\rm m})$. 
Since in the strong-coupling regime the condensate fraction $n_0/n_{\rm B}$ depends only weakly on the mass ratio, the dominant mass dependence of $\Delta_0/E_{\rm F}$ originates from the factor $(1+1/\gamma_{\rm m})$, which decreases monotonically when $\gamma_{\rm m}$ increases.
To verify that the dependence of the condensate fraction on the mass ratio plays indeed a minor role in $\Delta_0$,  Fig.~\ref{fig11} also reports $\Delta_0^2/E_{\rm F}^2$ with $n_0$ kept at its value for $\gamma_{\rm m} =1$  even when $\gamma_{\rm m}\neq 1$ (dotted lines). One sees that this quantity remains close to $\Delta_0^2/E_{\rm F}^2$ calculated with the appropriate value of $n_0$ for given $\gamma_{\rm m}$.

\begin{figure}[t]
    \centering
    \includegraphics[width=0.95\linewidth]{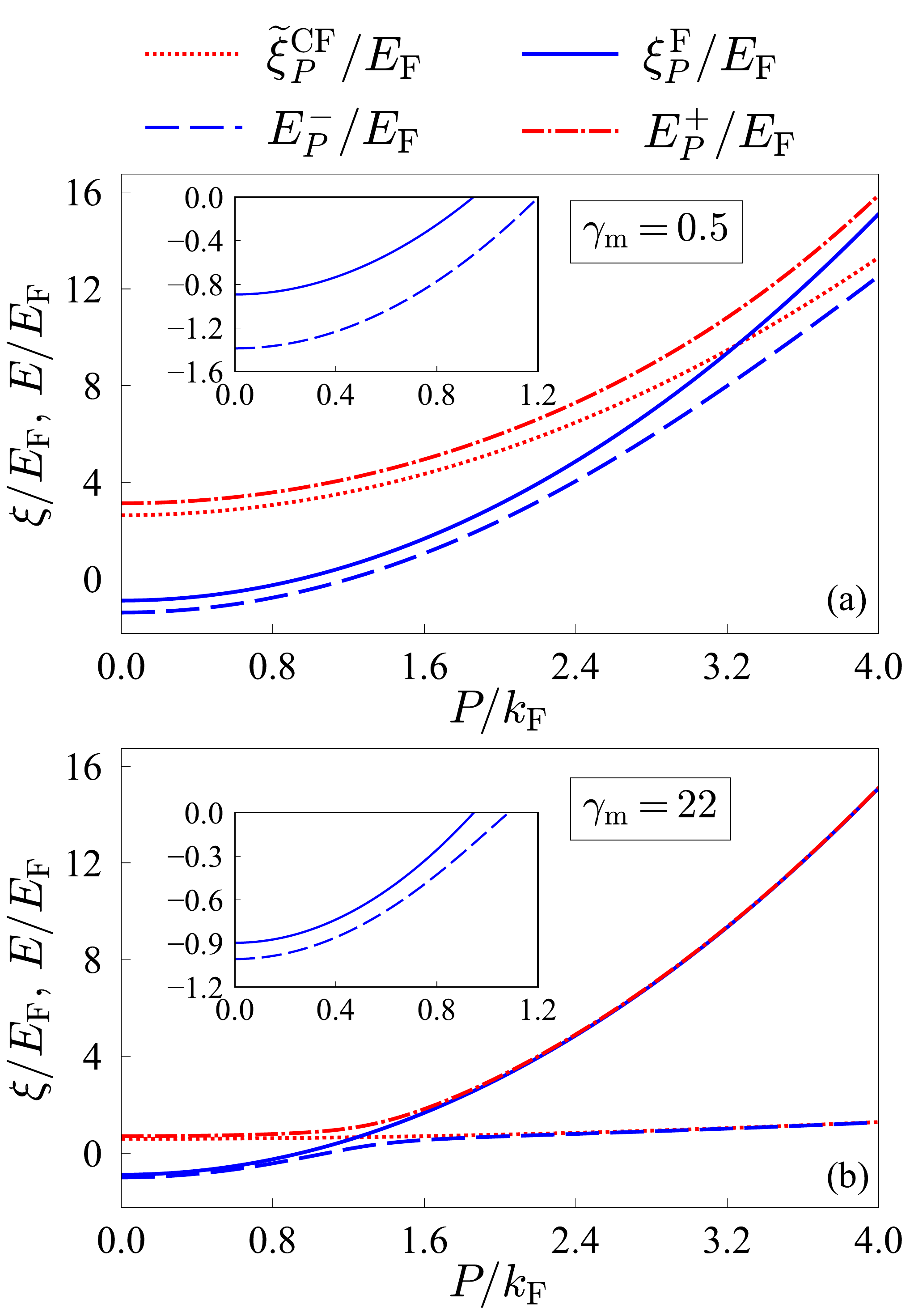}
    
    \caption{Dispersions $E_P^\pm$  of the two poles of the $T$-matrix and molecular and atomic dispersions $\widetilde{\xi}_P^{\;\mathrm{CF}}$ and $\xi_P^\mathrm{F}$ (in units of $E_\mathrm{F}$) as functions of $P/k_\mathrm{F}$. Data are obtained for a strong BF attraction $g = 3$, boson concentration $x = 0.175$ and mass ratios $\gamma_\mathrm{m} = 0.5$ (a), $\gamma_\mathrm{m} = 22$ (b). Insets: zoom on $E_P^-$ and $\xi_P^\mathrm{F}$ in the momentum region where $E_P^-<0$.}
    \label{fig12}
\end{figure}

Figure\til\ref{fig12} shows the dispersions of the two poles of the $T$-matrix for a fixed boson concentration ($x = 0.175$). The results show the characteristic level crossing (solid and dotted lines) and avoided crossing (dashed and dot-dashed lines) of a non-hybridized and hybridized two-level system, respectively. By comparing the two panels of Fig.~\ref{fig12}, one sees how increasing the mass ratio reduces the hybridization between the atomic and molecular branches, bringing the hybridized dispersions closer to the corresponding non-hybridized counterparts (see the insets for a zoom-in of the dispersions at negative energies, corresponding to occupied states of the composite fermions).

\section{Conclusions}\label{conclusions}

We have extended the diagrammatic $T$-matrix description of boson-fermion pairing and condensation in two-dimensional Bose-Fermi mixtures developed in Ref.~\cite{2dbfmix} to the experimentally relevant case of unequal boson and fermion masses.

Our results show that mass imbalance acts as an additional effective control parameter. While the chemical potentials and Tan's contact retain the same overall behavior observed in the mass-balanced case, the bosonic sector exhibits qualitatively new features. In particular, increasing the boson mass enhances the finite-momentum peak predicted in the boson momentum distribution, making it substantially more robust against the effects of a finite boson-boson repulsion. As a consequence, mixtures with sufficiently large mass ratios appear as especially promising candidates for the experimental observation of this feature.

Mass imbalance also affects the condensate properties. The near-universal dependence of the condensate fraction on the boson concentration, previously identified for equal masses, is preserved and becomes increasingly accurate for larger boson masses. At the same time, the strong-coupling molecular regime is modified through a reduction of the hybridization between atomic and molecular branches. Large mass ratios suppress the corresponding hybridization energy scale and drive the dressed molecular dispersions closer to their non-hybridized form.

On the theoretical side, we have shown analytically that the present $T$-matrix formalism reproduces the exact second-order perturbative expansion of the chemical potentials in the weak-coupling limit for arbitrary mass ratio and concentration, and we have verified this result numerically over a broad parameter range. This provides an important consistency check on the diagrammatic approximation and further supports its use across the crossover from weak to strong boson-fermion attraction.

The broad range of mass ratios available in ultracold heteronuclear mixtures, together with the recent realization of tunable Bose-Fermi platforms in transition-metal dichalcogenide heterostructures, makes the present results directly relevant to ongoing experimental efforts. More generally, our findings highlight the role of mass imbalance as a powerful tool for engineering pairing correlations, condensation, and molecular properties in low-dimensional Bose-Fermi systems. 

All the numerical data necessary to reproduce Figs.\til\ref{fig4}-\ref{fig12} and Fig.~\ref{fig15} are available online \cite{DatasetZenodo}.

\begin{acknowledgments}
We acknowledge the use of the parallel computing cluster of the Open Physics Hub at the Department of Physics and Astronomy of the University of Bologna. P.P.~acknowledges financial support from MUR project PE0000023-NQSTI (Italy) financed by the European Union - Next Generation EU and from European Union - NextGeneration EU through MUR under PNRR - M4C2 - I1.4 Contract No. CN00000013.
\end{acknowledgments}

\appendix

\section{Weak-coupling expansion}\label{appwc}

We explicitly derive the weak-coupling expansion of the diagrams selected in our $T$-matrix approach, and show that these recover the perturbative expansion of $\mu_\mathrm{B}$ and $\mu_\mathrm{F}$: 
    \begin{align}
        \frac{\mu_\mathrm{F}}{E_\mathrm{F}} & =1 + \,x \frac{1+\gamma_\mathrm{m}}{2\,\gamma_\mathrm{m}}\frac{1}{g} \nonumber \\
        &+x \frac{1+\gamma_\mathrm{m}}{4\,\gamma_\mathrm{m}} \left[\ln\left(\frac{\gamma_\mathrm{m}}{(1+\gamma_\mathrm{m})^2}\right)+\frac{\gamma_\mathrm{m}+1}{\gamma_\mathrm{m}-1}\ln \gamma_\mathrm{m}\right]\frac{1}{g^2} \label{mufpertfor} \\
         \frac{\mu_\mathrm{B}}{E_\mathrm{F}} &= \frac{1+\gamma_\mathrm{m}} {2\gamma_\mathrm{m}}\left[\frac{1}{g}+\frac{1}{g^2}\left(\frac{\gamma_\mathrm{m}\ln(\gamma_\mathrm{m})}{\gamma_\mathrm{m}-1}-\ln(\gamma_\mathrm{m}+1)-\frac{1}{2}\right)\right] \label{chempotfor}
\end{align}
obtained in \cite{montecarlo}.

\begin{figure*}[t]
    \centering
    \includegraphics[width=0.8\linewidth]{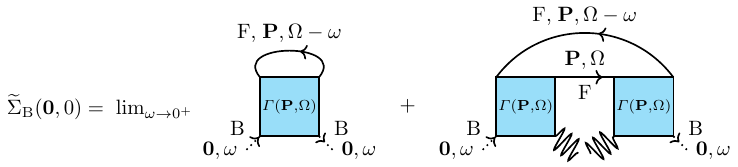}
    \caption{Feynman diagrams yielding the perturbative contributions to the boson chemical potential at first and second order in $1/g$, arising from the momentum integration outside of $\left[P_{T_-},P_{T_+}\right]$. The dotted lines with arrows indicate the momentum flow.}
    \label{fig13}
\end{figure*}

\subsection{The \texorpdfstring{$\varGamma$-matrix}{Gamma}  }

By expanding $T_2(\textbf{P},\Omega)$  in powers of $1/g$, it is easy to identify the following perturbative orders of the $\varGamma$-matrix $\varGamma(\textbf{P},\Omega) = \varGamma^{(1)}+ \varGamma^{(2)}(\textbf{P},\Omega)+~\hdots$
\begin{align}
\varGamma^{(1)}
&=
\frac{\pi}{g\,m_{\mathrm r}},
\label{gamma(1)}
\\
\varGamma^{(2)}(\mathbf P,\Omega)
&=
\frac{\pi}{g^{2}m_{\mathrm r}}
\bigg[
\frac12
\ln\left(\frac{{\textbf{P}^2}/{(2M)}-\mu_\mathrm{F}-\mu_\mathrm{B}-i\Omega}{k_\mathrm{F}^2/(2m_\mathrm{r})}\right)\nonumber\\
&\hspace{2cm}+\frac{\pi}{m_{\mathrm r}}I_{\mathrm F}(\mathbf P,\Omega)\bigg]
\label{gamma(2)}
\end{align}

or, in integral form,
\begin{align}
\varGamma^{(2)}(\mathbf P,\Omega)
&=
\frac{\pi^{2}}{g^{2}m_{\mathrm r}^{2}}
\int\frac{\dd^2 q}{(2\pi)^2}
\Biggl[
\frac{
\Theta(-\xi^{\mathrm F}_{\mathbf P-\mathbf q})-1
}{
\xi^{\mathrm F}_{\mathbf P-\mathbf q}
+\xi^{\mathrm B}_{\mathbf q}
-i\Omega
}
\nonumber\\
&\qquad
+\frac{1}{
\mathbf q^{2}/(2m_{\mathrm r})
+k_\mathrm{F}^2/(2m_\mathrm{r})
}
\Biggr].\label{gamma(2)int}
\end{align}

Note that the first order is independent of the (2+1)-momentum $(\textbf{P},\Omega)$.

\subsection{The boson chemical potential}
According to the Hugenholtz-Pines theorem \eqref{hp}, in the absence of BB interactions, the integral that determines the boson chemical potential is 
\begin{equation}
\Sigma_\mathrm{B}(\mathbf{0},0)
=
\int \frac{\dd^2 P\,\dd\Omega}{(2\pi)^3}
\,T(\mathbf{P},\Omega)\,
G_\mathrm{F}^0(\mathbf{P},\Omega)\,
\e^{i\Omega 0^+}.
\label{a1}
\end{equation}
We define the momenta $P_{T_\pm}$ as the solutions of the equations 
\begin{equation}
    P = \lim_{\Omega \to 0^+}\sqrt{2m_\mathrm{F}\left[\mu_\mathrm{F}\pm n_0 |\varGamma(P,\Omega)|\right]}
\end{equation}
respectively. $P_{T_+}$ represents the momentum of the Fermi step introduced by the $T$-matrix, and it provides a natural cutoff for the momentum integral of Eq.~\eqref{a1}. $P_{T_-}$ sets the lower bound of the range $\left[P_{T_-},P_{T_+}\right]$ where the $T$-matrix cannot be represented as a convergent geometric series uniformly in $P,\,\Omega$ since 
\begin{equation}
    T(\textbf{P},\Omega) \not = \varGamma(\textbf{P},\Omega) \,\sum_{n = 0}^\infty\left[n_0 G_\mathrm{F}^0 (\textbf{P},\Omega)\, \varGamma(\textbf{P},\Omega)\right]^n \end{equation}
    for $\Omega$ small enough when $ P \in \left[P_{T_-},P_{T_+}\right]$. At leading order,
    \begin{equation}
        P_{T_{\pm}} = \sqrt{2m_\mathrm{F}\left(\mu_\mathrm{F}\mp \mu_\mathrm{F}^{(1)}\right)},
    \end{equation}
where $\mu_\mathrm{F}^{(1)} = E_\mathrm{F}\,x {(\gamma_\mathrm{m}+1)}/{(2 g \gamma_\mathrm{m})}$ is the contribution of order O$\left(1/g\right)$ to the fermion chemical potential appearing in Eq.~\eqref{mufpertfor} and derived in Sec.~\ref{sec:anchempot} below.

To identify the different perturbative orders of $\mu_\mathrm{B}$, it is useful to split the integration domain $\left(0,P_{T_+}\right) = \left(0,P_{T_-}\right]\cup \left(P_{T_-},P_{T_+}\right)$. By restricting the momentum interval to $0<P<P_{T_-}$, we can identify the first contributions of order less than or equal to $\mathrm{O}\left(1/g^2\right)$ from the diagrammatic expansion in Fig.~\ref{fig13}, which is obtained by approximating the $T$-matrix by the first two terms of the convergent geometric series, and we denote the corresponding self-energy by $\widetilde{\Sigma}_\mathrm{B}(\textbf{0},0)$.

By inserting the expansion of the $\varGamma$-matrix given by Eqs.~\eqref{gamma(1)}-\eqref{gamma(2)int}, and of the fermion chemical potential given by Eq.~\eqref{mufpertfor} -- while it is sufficient to take $\mu_\mathrm{B} \to 0^-$ inside the $\varGamma$ and $n_0 = n_\mathrm{B}$, noting from Eq.~\eqref{n0pert} that the leading correction to the boson condensate density is of order O$(1/g^2)$ -- it is easy to find
\begin{align}
    &\frac{\widetilde{\Sigma}_\mathrm{B}(\textbf{0},0)}{E_\mathrm{F}} = \frac{1+\gamma_\mathrm{m}}{2\,\gamma_\mathrm{m}}\frac{1}{g}+\frac{1+\gamma_\mathrm{m}}{4\gamma_\mathrm{m}}\bigg(\frac{2\gamma_\mathrm{m}\ln(\gamma_\mathrm{m})}{\gamma_\mathrm{m}-1}\nonumber\\
    &-2\ln(\gamma_\mathrm{m}+1)-1+2x \frac{\gamma_\mathrm{m}+1}{\gamma_\mathrm{m}}\bigg)\frac{1}{g^2},\label{chempotfortilde}
\end{align}
valid up to corrections of order higher than O$\left(1/g^2\right)$. 

The concentration-dependent term in the last line of Eq.~\eqref{chempotfortilde} comes from the perturbative expansion of the fermion chemical potential, and is therefore a consequence of treating fermions in the grand-canonical ensemble. Such a term is canceled by the remaining momentum integral over the interval $(P_{T_-},P_{T_+})$, given by 
\begin{eqnarray}
    \overline{\Sigma}_\mathrm{B}(\mathbf{0},0) &=& \int_{P_{T_-}}^{P_{T_+}} \frac{\dd P}{2\pi}\,P\int \frac{\dd \Omega}{2\pi}\, T(\textbf{P},\Omega)G_\mathrm{F}^0(\textbf{P},\Omega)\e^{i\Omega 0^+}\nonumber\\ 
    &=& -E_\mathrm{F}\frac{x}{2}\left(\frac{\gamma_\mathrm{m}+1}{\gamma_\mathrm{m}}\right)^2\frac{1}{g^2},\label{correction}
\end{eqnarray} 
which is valid up to corrections of order higher than O$\left(1/g^2\right)$. We stress again that  $\overline{\Sigma}_\mathrm{B}(\mathbf{0},0)$ comes from the region in momentum space where the formal sum \eqref{tmatrixdef} defining the $T$-matrix differs from a convergent geometric series, and is therefore a consequence of the way we formally accounted for an infinite number of condensate insertions. 
Therefore, summing up the two contributions \eqref{chempotfortilde}, \eqref{correction} into $\mu_\mathrm{B} = \widetilde{\Sigma}_\mathrm{B}(\textbf{0},0)+\overline{\Sigma}_\mathrm{B}(\textbf{0},0)$ leaves no dependence on $x$ and yields the result of Eq.~\eqref{chempotfor}. 

Interestingly, the same perturbative expansion is recovered by representing the $T$-matrix as a geometric series in the whole (2+1)-momentum space, provided that the zero-temperature limit is performed after the frequency sum: up to corrections of order higher than O$\left(1/g^2\right)$,
\begin{align} 
  &  \int \frac{\dd ^2 P\,\dd \Omega}{(2\pi)^3}\,T(\textbf{P},\Omega)G_\mathrm{F}^0(\textbf{P},\Omega)\e^{i\Omega 0^+}\nonumber \\
  &= \int \frac{\dd ^2 P}{(2\pi)^2} \lim_{\beta \to \infty} \frac{1}{\beta}\sum_{\Omega_n} \bigg(\varGamma(\textbf{P},\Omega_n)\nonumber\\
  &\hspace{0.2cm}+ n_0\varGamma(\textbf{P},\Omega_n)^2 G_\mathrm{F}^0(\textbf{P},\Omega_n)\bigg)G_\mathrm{F}^0(\textbf{P},\Omega_n)\e^{i\Omega_n 0^+},\label{reg}
\end{align}
with $\Omega_n = (2n+1)\pi/\beta,\,n \in \mathbb{Z}$ being the fermion Matsubara frequencies. The order between the sum and the limit is important to make sense of the singular behavior of $G_\mathrm{F}^0(\mathbf{P},\Omega)^2$ at $i\Omega = \xi_\mathbf{P}^\mathrm{F}$ on the right-hand side of Eq.~\eqref{reg}. In particular, $\lim_{\beta \to \infty} \beta^{-1}\sum_{\Omega_n}G_\mathrm{F}^0(\textbf{P},\Omega_n)^2\e^{i\Omega_n0^+} = -\delta\left(\xi_\textbf{P}^\mathrm{F}\right)$, and the second line on the right-hand side of Eq.~\eqref{reg} yields a contribution proportional to $x$ that cancels the dependence on the concentration originating from the expansion of $\mu_\mathrm{F}$ in the first line. The frequency integrand on the left-hand side instead does not exhibit any double pole singularity, and therefore we can take the continuum limit of $\Omega$ from the beginning. As seen above, the analogous cancellation between terms dependent on the boson concentration occurs automatically.

\subsection{The fermion chemical potential}\label{sec:anchempot}

\begin{figure*}
    \centering
    \includegraphics[width=1.0\linewidth]{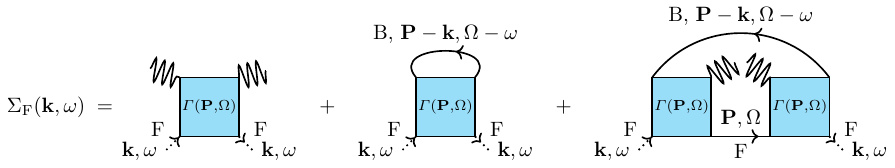}
    \caption{Feynman diagrams yielding the perturbative contributions to the fermion self-energy at first and second order in $1/g$. The dotted lines with arrows indicate the momentum flow.}
    \label{fig14}
\end{figure*}
Let us now adapt the arguments developed for the boson chemical potential to the fermion case. According to the Eq.~\eqref{nf}, the integral that determines the fermion density reads
    \begin{equation} 
        n_\mathrm{F} = \int \frac{\dd^2 k\,\dd \omega}{(2\pi)^3}\, \frac{G_\mathrm{F}^0(\textbf{k},\omega)}{1-G_\mathrm{F}^0(\textbf{k},\omega)\Sigma_\mathrm{F}(\textbf{k},\omega)}\e^{i\omega 0^+}. \label{neffapp}
        \end{equation}
We define the momenta $k_{\pm}$ as the solutions to the equations 
\begin{equation}
    k = \sqrt{2m_\mathrm{F}\left[\mu_\mathrm{F}\pm\left|\Sigma_\mathrm{F}^{\mathrm{max}}\left(k\right)\right|\right]} 
\end{equation}
respectively. Here, $|\Sigma^\mathrm{max}_\mathrm{F}(k)|$ denotes the maximum value of $|\Sigma_\mathrm{F}(k,\omega)|$ as a function of $\omega$, at fixed $k$. $k_\pm$ set the bounds of the range $\left[k_-,k_{+}\right]$ where $\left(1-G_\mathrm{F}^0(\textbf{k},\omega)\Sigma_\mathrm{F}(\textbf{k},\omega)\right)^{-1}$ cannot be represented as a convergent geometric series uniformly in $k,\,\omega$: 

\begin{align}
    \left[1-G_\mathrm{F}^0(\textbf{k},\omega)\Sigma_\mathrm{F}(\textbf{k},\omega)\right]^{-1}\neq~\sum_{n = 0}^\infty\left[G_\mathrm{F}^0 (\textbf{k},\omega)\, \Sigma_\mathrm{F}(\textbf{k},\omega)\right]^n\end{align}  
for $\omega$ small enough when $k \in [k_-,k_+]$. For notational convenience, we define $\Sigma_\mathrm{F}^\pm = |\Sigma_\mathrm{F}^\mathrm{max}(k_\pm)|$. It can be shown that, at leading order, $k_\pm = P_{T_\pm}$.

It is useful to split the integration domain $(0,\infty) = (0,k_-)\cup[k_-,k_+]\cup(k_+,\infty)$ and hence write Eq.~\eqref{neffapp} as
\begin{align}
    &n_\mathrm{F} = \frac{k_{-}^2}{4\pi} + I_1 + I_2,\label{nfapp}
\end{align}  
with
\begin{align}
&I_1 = \int'\frac{\dd k}{2\pi}\,k \int \frac{\dd \omega}{2 \pi}\,G_\mathrm{F}^0(\textbf{k},\omega)\bigg(G_\mathrm{F}^0(\textbf{k},\omega)\Sigma_\mathrm{F}(\textbf{k},\omega)\nonumber\\
&\hspace{2cm}+G_\mathrm{F}^0(\textbf{k},\omega)^2 \Sigma_\mathrm{F}(\textbf{k},\omega)^2\bigg)\e^{i\omega 0^+},\label{i1}\\
&I_2 = \int_{k_-}^{k_+}\frac{\dd k}{2\pi}\,k \int \frac{\dd \omega}{2 \pi}\,\frac{G_\mathrm{F}^0(\textbf{k},\omega)}{1-G_\mathrm{F}^0(\textbf{k},\omega)\Sigma_\mathrm{F}(\textbf{k},\omega)} \e^{i\omega 0^+}\label{i2}
\end{align}
valid up to higher order corrections in $\Sigma_\mathrm{F}(\textbf{k},\omega)$. In the above, $\int' \dd k\,$ denotes the integral over $(0,\infty)\setminus [k_-,k_+]$.

Using $k_-^2 = 2m_\mathrm{F}(\mu_\mathrm{F}-\Sigma_\mathrm{F}^-)$, Eq.~\eqref{nfapp} can be recast into the following equation for the fermion chemical potential:
\begin{align}
    \mu_\mathrm{F} = E_\mathrm{F}-\frac{2\pi}{m_\mathrm{F}}I_1-\left(\frac{2\pi}{m_\mathrm{F}}I_2-\Sigma_\mathrm{F}^-\right).
\end{align}

To identify the different perturbative orders of $\mu_\mathrm{F}$,  we begin from the fermion self-energy
\begin{align}
    \Sigma_\mathrm{F}(\mathbf{k},\omega) = -\int \frac{\dd ^2 P\,\dd \Omega}{(2\pi)^3}\,T(\textbf{P},\Omega)G_\mathrm{B}^0(\textbf{P}-\mathbf{k},\Omega-\omega)\e^{i\Omega 0^+} \! ,\label{a2}
\end{align}
where we have dropped the term $n_0 \varGamma(\textbf{k},\omega)$ since here we want to focus on the expansion of the $T$-matrix. Carrying out the steps discussed in the previous section, one deduces the contribution to $\Sigma_\mathrm{F}(\textbf{k},\omega)$ from the momentum interval where the convergence of the geometric series for $T(\textbf{P},\Omega)$ is spoiled:
\begin{align}
    \overline{\Sigma}_\mathrm{F}&(\textbf{k},\omega) = -\int_{P_{T_-}}^{P_{T_+}} \frac{\dd P}{2\pi}\, P\nonumber\\
    \times &\int_0^{2\pi} \frac{\dd \theta}{2\pi}\int \frac{\dd \Omega}{2\pi}\, T(\textbf{P},\Omega)G_\mathrm{B}^0(\textbf{P}-\mathbf{k},\Omega-\omega)\e^{i\Omega 0^+}\label{nogeomser},
    \end{align}
where $\theta$ is the angle between $\textbf{P}$ and $\textbf{k}.$ By contour integration, the second line of Eq.~\eqref{nogeomser} is found to be of order O$(1/g^2)$. In turn, this implies that $ \overline{\Sigma}_\mathrm{F}(\textbf{k},\omega)$ is of order higher than O$\left(1/g^2\right)$ and that (at second-order) $\Sigma_\mathrm{F}(\textbf{k},\omega)$ can be evaluated by treating the $T$-matrix in Eq.~\eqref{a2} as a convergent geometric series. 

We can identify the contributions of order less than or equal to O$(1/g^2)$ from the diagrammatic expression in Fig.~\ref{fig14}. The contribution to O$(1/g)$ is independent of $\textbf{k},\omega$, and reads $\Sigma_\mathrm{F}^{(1)} = n_\mathrm{B}\varGamma^{(1)}$. Plugging this and the second order $\Sigma_\mathrm{F}^{(2)}(\textbf{k},\omega)$ in Eq.~\eqref{i1}, a direct calculation yields  $I_1 = 0$ up to corrections of order higher than O$(1/g^2)$. Eq.~\eqref{i2} instead yields  
\begin{equation}
       I_2= -\frac{m_\mathrm{F}}{2\pi}\left( \lim_{\omega \to 0^+}\Sigma_\mathrm{F}\left(\widetilde{k}_\mathrm{F},\omega\right) -\Sigma_\mathrm{F}^-\right),\label{i2fin}
\end{equation}
to second order, where $\widetilde{k}_\mathrm{F}$ is the Fermi step momentum of the fermion momentum distribution $n_\mathrm{F}(\textbf{k})$. Actually, since $\widetilde{k}_\mathrm{F}$ and the non-interacting Fermi momentum $k_\mathrm{F}$ differ by a term of order $(1/g)$, up to higher order corrections we can replace $\widetilde{k}_\mathrm{F}$ by $k_\mathrm{F}$ in Eq.~\eqref{i2fin}.

In summary, we are left with 
\begin{equation}
    \mu_\mathrm{F} = E_\mathrm{F} + \lim_{\omega \to 0^+}\Sigma_\mathrm{F}(k_\mathrm{F},\omega), \label{luttinger}
\end{equation}
which is the Luttinger theorem for Fermi liquids \cite{Luttinger-1960}. 
The present proof thus shows that our choice of Feynman diagrams respects the Luttinger theorem in the weak-coupling limit.

The perturbative result \eqref{mufpertfor} for $\mu_{\rm F}$ is finally obtained by expanding $\Sigma_\mathrm{F}(k_\mathrm{F},\omega\to 0^+)$ to second order in $1/g$. Again, the first- and second-order terms in $1/g$ can be identified from the diagrammatic expression in Fig.~\ref{fig14}. 

\subsection{Numerical check}
\begin{figure}[htbp]
    \centering
        \includegraphics[width=1.0\linewidth]{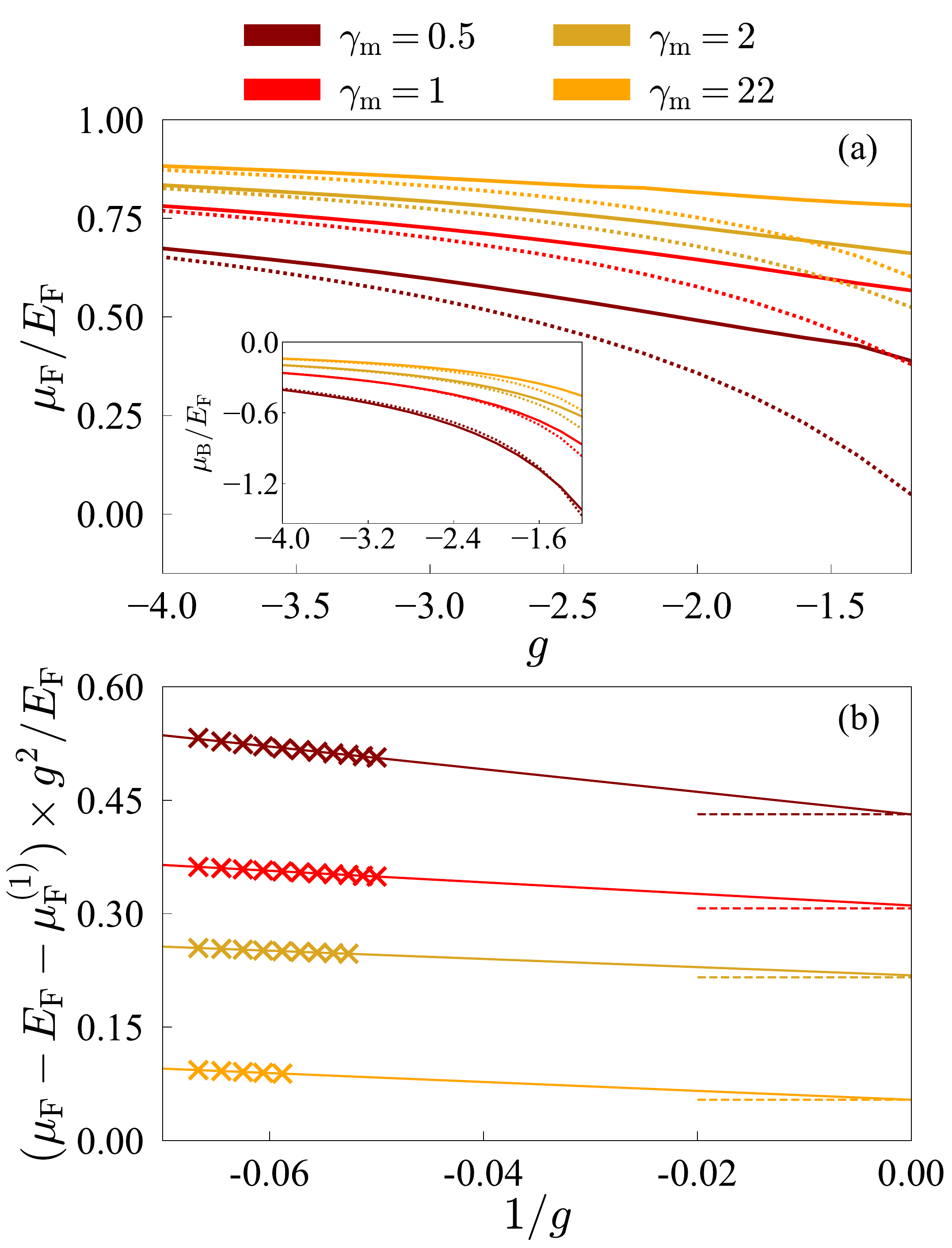}
         \caption{(a): Fermion chemical potential $\mu_\mathrm{F}$, in units of $E_\mathrm{F}$, as a function of the BF coupling $g$, for $x = 1$ and several values of mass ratio $\gamma_{\rm m}$. The dotted lines represent the weak-coupling result Eq.~\eqref{mufpertfor}. Inset: corresponding boson chemical potential (full lines) and its perturbative expression Eq.~\eqref{chempotfor} (dotted lines). (b): the crosses represent the quantity $(\mu_\mathrm{F}-E_\mathrm{F}-\mu_\mathrm{F}^{(1)})g^2/E_\mathrm{F}$, as a function of $1/g$, for $x = 1$; the full lines represent the fit to the model $f(g) =A + B/g$; the dashed lines represent the expected asymptotic value $\mu_\mathrm{F}^{(2)}g^2$ of the fitting curve in the limit $1/g \to 0^-$.}
    \label{fig15}
\end{figure}

Finally, we provide a numerical check of the results for the chemical potentials obtained by the full numerical solution of the set of equations \eqref{nf}-\eqref{hp} against their weak-coupling benchmarks. 
Fig.~\ref{fig15}(a) shows that, in the weak coupling regime $g \lesssim -2$, the numerical results for $\mu_{\rm F}$ approach the asymptotic form given by \eqref{mufpertfor}. 
To isolate the higher-order contributions, we subtract the zeroth- and first-order terms, given by $E_\mathrm{F}+\mu_\mathrm{F}^{(1)}=E_\mathrm{F}\big(1+x {(\gamma_\mathrm{m}+1)}/{(2 g \gamma_\mathrm{m})}\big)$, multiply the residual by $g^2$, and fit the resulting quantity to $f(g) = A + B/g$, in the BF coupling range $-20 \le g \le -15$. 

Fig.\til\ref{fig15}(b) shows the best-fit curve for several values of mass ratio for the selected case $x = 1$. 
The results of the fitting procedure demonstrate an excellent agreement between numerical data and the asymptotic result obtained from the perturbative expansion Eq.~\eqref{mufpertfor}, thus confirming its non-linear dependence on the mass ratio. 

The inset of Fig.~\ref{fig15}(a) shows that the numerical curve for the boson chemical potential approaches the asymptotic form given by Eq.~\eqref{chempotfor}. We have applied the same procedure used to verify the second-order coefficient of $\mu_\mathrm{F}$ to the case of $\mu_\mathrm{B}$, confirming the validity of the expansion in Eq.~\eqref{chempotfor} at second-order. In addition, for both the boson and the fermion chemical potentials, we have observed a similar good agreement between the numerical and the analytical results also for $x < 1$.

In \cite{montecarlo}, Eqs.~\eqref{mufpertfor} and \eqref{chempotfor} were tested against independent fixed-node diffusion Quantum Monte Carlo simulations, providing further evidence that they represent the correct second-order perturbative approximation for the thermodynamic quantities of the mixture. Therefore, the present analysis validates our choice of Feynman diagrams in the weak-coupling regime, up to corrections of order higher than O$(1/g^2)$.

\section{Strong-coupling expressions}\label{asssc}
Here, we provide closed-form expressions for the strong-coupling set of equations \eqref{nfsc}-\eqref{hpsc}. The momentum integral in Eq.~\eqref{nbsc} can be performed analytically, yielding
\begin{align}
&n_\mathrm{B}-n_0 = \int \frac{\dd^2 P}{(2\pi)^2}\,\left[ u_\textbf{P}^2 \Theta(-E_\textbf{P}^+)+v_\textbf{P}^2 \Theta(-E_\textbf{P}^-) \right]\nonumber\\
    &=\frac{n_\mathrm{F}}{4E_\mathrm{F}}\bigg\{ {\widetilde{\mu}}_\mathrm{CF}(1+\gamma_\mathrm{m}) +  {\mu}_\mathrm{F} +  {\mathcal{K}}-2\frac{1 + \gamma_\mathrm{m}}{\gamma_\mathrm{m}}\nonumber\\
    &\times\bigg[ \sqrt{4  {\Delta}_0^2 + \left( {\widetilde{\mu}}_\mathrm{CF}- {\mu}_\mathrm{F}\right)^2}-\sqrt{\left(4  {\Delta}_0^2+\frac{{\mathcal{J}}^2}{4(1+\gamma_\mathrm{m})^2} \right)}\bigg]\bigg\}\label{NCFSTRONG}
\end{align}
with
\begin{align}
     & {\mathcal{K}} = \sqrt{4  {\Delta}_0^2 (1+\gamma_\mathrm{m}) + \left( {\widetilde{\mu}}_\mathrm{CF}(1+\gamma_\mathrm{m}) -  {\mu}_\mathrm{F}\right)^2},\\
    &  {\mathcal{J}} =  {\widetilde{\mu}}_\mathrm{CF}(\gamma_\mathrm{m}^2 + 3\gamma_\mathrm{m}+2) -  {\mu}_\mathrm{F}(2+\gamma_\mathrm{m})+\gamma_\mathrm{m} {\mathcal{K}}.
\end{align} 
The momentum integral in Eq.~\eqref{hpsc} instead  yields
\begin{align}
     \mu_\mathrm{B} &=-\epsilon_0\left(\frac{1+\gamma_\mathrm{m}}{\gamma_\mathrm{m}}\right)^2\bigg[ \ln \left(2(1+\gamma_\mathrm{m})\right)\nonumber\\
     &+\ln\left({\mu}_\mathrm{F} -{\widetilde{\mu}}_\mathrm{CF} +\sqrt{4{\Delta}_0^2+\left({\widetilde{\mu}}_\mathrm{CF} -{\mu}_\mathrm{F} \right)^2}\right)\nonumber\\
    &-\ln\bigg(2\sqrt{4{\Delta}_0^2(1+\gamma_\mathrm{m})^2+\frac{1}{4} {\mathcal{J}}^2}-{\mathcal{J}}\bigg)\bigg].\label{c17}
\end{align}

Finally, the fermion momentum distribution appearing in Eq.~\eqref{nfsc} is given by
\begin{equation}
n_{\rm F}({\bf k})=\sum_{z_i} \lim_{z \to z_i}(z-z_i)\,G_\mathrm{F}(\textbf{k},z)\,\Theta(-z_i) ,
\end{equation}
where 
\begin{align}
     G_\mathrm{F}(\textbf{k},z)^{-1} = z - \xi_\textbf{k}^\mathrm{F} -\frac{\Delta_0^2}{z - \widetilde{\xi}_\textbf{k}^{\;\mathrm{CF}}}-  \frac{{2 \pi \epsilon_0 n_\mathrm{CF}}/{m_\mathrm{r}}}{z + \xi_\textbf{k}^\mathrm{B}},
\end{align}
$n_\mathrm{CF}$ is the right-hand side of Eq.~\eqref{NCFSTRONG}, and $z_i$ are the three (real) solutions of the equation $G_\mathrm{F}(\textbf{k},z)^{-1} = 0$.
The momentum integral in Eq.~\eqref{nfsc}  remains to be evaluated numerically.

\bibliography{BFmimb2dRE}

\end{document}